\documentclass[12pt,preprint]{aastex}
\usepackage{subfigure}
\begin{document}

\title{Evidence for Evolution Among Primordial Disks in the 5 Myr Old Upper Scorpius OB Association}

\author{S. E. Dahm\altaffilmark{1,2}}

\altaffiltext{1}{W. M. Keck Observatory, 65-1120 Mamalahoa Hwy, Kamuela, HI 96743}
\altaffiltext{2}{Visiting Astronomer at the Infrared Telescope Facility, which is operated by the University of Hawaii under Cooperative Agreement no. NNX-08AE38A with the National Aeronautics and Space Administration, Science Mission Directorate, Planetary Astronomy Program.}

\begin{abstract}
Moderate-resolution, near-infrared spectra between 0.8 and 5.2 $\mu$m were obtained for 12 late-type 
(K0--M3) disk-bearing members of the $\sim$5 Myr old Upper Scorpius OB association using SpeX 
on the NASA Infrared Telescope Facility. For most sources, continuum excess emission first becomes 
apparent between $\sim$2.2 and 4.5 $\mu$m and is consistent with that produced by single-temperature 
blackbodies having characteristic temperatures ranging from $\sim$500 to 1300 K. The near-infrared 
spectra for 5 of 12 Upper Scorpius sources exhibit Pa$\gamma$, Pa$\beta$ and Br$\gamma$ emission, 
indicators of disk accretion. Using a correlation between Pa$\beta$ and Br$\gamma$ emission line 
luminosity and accretion luminosity, mass accretion rates ($\dot{M}$) are derived for these sources 
that range from $\dot{M} = 3.5\times10^{-10}$ to 1.5$\times10^{-8}$ M$_{\odot}$ yr$^{-1}$. Merging 
the SpeX observations with {\it Spitzer Space Telescope} mid-infrared (5.4--37.0 $\mu$m) spectroscopy 
and 24 and 70 $\mu$m broadband photometry, the observed spectral energy distributions are compared 
with those predicted by two-dimensional, radiative transfer accretion disk models. Of the 9 Upper 
Scorpius sources examined in this analysis, 3 exhibit spectral energy distributions that are most consistent with 
models having inner disk radii that substantially exceed their respective dust sublimation radii. 
The remaining Upper Scorpius members possess spectral energy distributions that either show significant 
dispersion among predicted inner disk radii or are best described by models having inner disk rims 
coincident with the dust sublimation radius.
\end{abstract}


\section{Introduction}

A fundamental consequence of protostellar collapse is the formation of a viscous accretion disk that 
transfers gas onto the stellar photosphere. Such disks are the progenitors of planetary systems and 
their subsequent evolution and ultimate dispersal have extraordinary implications for the formation 
of planets, orbital dynamics of planetary mass bodies, and the processing of dust grains and volatiles 
within the disk. The timescale of disk dissipation within the terrestrial region has been reasonably 
well-established by ground-based (Haisch et al. 2001; Mamajek et al. 2004) and {\it Spitzer Space
Telescope} (Uchida et al. 2004; Silverstone et al. 2006) observations to be $\le$10 Myr. By inference, 
planetary systems must be in an advanced evolutionary stage within this period, before significant 
quantities of gas and dust are depleted from the disk. Disk evolution is expected to proceed from the 
interior outward (Dullemond \& Dominik 2005), however, sub-millimeter and mid-infrared observations by
Cieza et al. (2008) suggest that inner disks only begin to dissipate after the outer disk has been 
significantly depleted of mass. 

Pre-main sequence stars exhibiting spectral energy distributions (SED) that are indicative of an 
optically thin disk interior surrounded by an optically thick outer disk are referred to as transition 
disk objects (Strom et al. 1989; Najita et al. 2007; Muzerolle et al. 2010). Such disks are suggestive 
of having experienced significant evolution from the continuous disk structures that are associated 
with classical T Tauri stars (CTTS). The 
duration of this transitional phase has been inferred from population statistics of star forming regions 
to be only of order $\sim$10$^{5}$ yr (Hartmann 2009; Luhman et al. 2010). Transition-like SEDs, 
however, can arise from various pathways and may not be representative of a direct evolutionary sequence 
from primordial disk to debris disk (Najita et al. 2007). Dust grain growth and mid-plane settling 
(Dullemond \& Dominik 2005), giant planet formation and dynamical clearing (Calvet et al. 2002), 
disk photoevaporation (Alexander et al. 2006), and the presence of a stellar companion (Ireland \& Kraus 2009)
have been suggested as disk clearing mechanisms capable of producing transition-like SEDs. 
An alternate evolutionary path from primordial disk to debris disk has been proposed by Currie et al. (2009),
who suggest that disks exhibiting reduced levels of near- and mid-infrared excess emission are
indicative of reduced masses of small dust grains at all disk radii. Such homologously depleted disks
counter the canonical inside-out disk evolutionary scenario.

The Upper Scorpius OB association is a critically important region for studies of disk evolution. At
$\sim$145 pc distant, it is among the nearest OB associations to the Sun (Blaauw 1991; de Zeeuw et al. 1999) 
and has a well-established age of $\sim$5 Myr (Preibisch \& Zinnecker 1999; Preibisch et al. 2002), when
most ($\sim$80\%) optically thick, protoplanetary disks have dissipated (Haisch et al. 2001; Hernandez 
et al. 2007). Carpenter et al. (2006) conducted a {\it Spitzer} 4.5--16.0 $\mu$m photometric survey 
of 218 confirmed association members. These sources were compiled from 
{\it Hipparcos} astrometry (de Zeeuw et al. 1999), color-magnitude diagrams and \ion{Li}{1} $\lambda$6708 
follow-up observations (Preibisch \& Zinnecker 1999; Preibisch et al. 2002), and X-ray detected late-type 
stars with \ion{Li}{1} $\lambda$6708 follow-up (Walter et al. 1994; Kunkel 1999; K\"{o}hler et al. 2000). 
Given that the membership selection criteria were based upon stellar properties unrelated to circumstellar 
material, the sample is considered to be unbiassed toward the presence or absence of disks. Carpenter et al. (2006) 
found that 24 of 127 (19\%) K- and M-type stars in their sample exhibit infrared excesses similar to 
Class II sources in the Taurus-Auriga star forming region. For comparison, Luhman et al. (2010) carried out a comparable {\it Spitzer}
IRAC and MIPS survey of 348 Taurus-Auriga members, finding that the disk fraction steadily declines 
from $\sim$75\% for solar mass stars to 45\% for low-mass stars and brown dwarfs (0.3--0.01 M$_{\odot}$).

To better characterize disk emission among the Upper Scorpius infrared excess stars identified by 
Carpenter et al. (2006), Dahm \& Carpenter (2009) examined {\it Spitzer} mid-infrared spectra for a 
substantial fraction (26 of 35) of these sources: 8 early-type (B+A) stars and 18 late-type (K+M) stars. 
In general, excess emission among the late-type Upper Scorpius population becomes apparent between 2.2 and 
4.5 $\mu$m, a region sampled only by Two-Micron All Sky Survey (2MASS) $K_{S}-$band photometry and the {\it Spitzer} 
Infrared Array Camera (IRAC) [4.5] fluxes. Compared to Class II sources in Taurus-Auriga, Dahm \& Carpenter (2009)
found the disk population in Upper Scorpius to exhibit reduced levels of near- and mid-infrared excess emission and
an order of magnitude lower mass accretion rates. The apparent abundance of depleted inner disk systems in 
Upper Scorpius relative to the number of such objects in Taurus-Auriga led Dahm \& Carpenter (2009) to suggest 
that such disks represent a common evolutionary pathway. Near-infrared spectra spanning the 2.2--4.5 $\mu$m 
region, however, were needed to complement the {\it Spitzer} IRAC and Infrared Spectrograph (IRS) observations 
and to isolate the onset of continuum excess emission from the terrestrial disk regions.  

To further examine emission arising from the depleted disk interiors of the Upper Scorpius sample, 0.8--5.2 $\mu$m 
moderate-resolution spectra for 12 of 18 late-type disk-bearing stars from Dahm \& Carpenter (2009) were obtained 
using SpeX on the NASA Infrared Telescope Facility (IRTF) on Mauna Kea. The SpeX spectroscopy permits the unambiguous 
detection of continuum excess emission arising from hot dust in the disk interiors of the Upper Scorpius sample. 
Merging these observations with {\it Spitzer} IRS spectra and Multi-band Imaging Photometer for {\it Spitzer} (MIPS) 
24 and 70 $\mu$m photometry from Carpenter et al. (2009), the resulting 0.8--70 $\mu$m SEDs are compared with the 
two-dimensional, radiative transfer accretion disk models of Robitaille et al. (2006). These models are used to 
constrain inner disk radii for the observed Upper Scorpius excess sources and to examine their SEDs for evidence
of inner disk evolution. 

In Section 2 the observed Upper Scorpius sample is described and placed into context with the greater late-type
stellar population of the association. Details of the SpeX observations and analysis are also provided. The
near-infrared continuum excess spectra and blackbody fits to these spectra are discussed in Section 3. Next
(Section 4) accretion luminosities and mass accretion rates are derived for suspected accretors using Pa$\beta$
and Br$\gamma$ emission line luminosities. The merged 0.8--70 $\mu$m SEDs are then compared in Section 5
with the accretion disk models of Robitaille et al. (2006). Constraints for inner disk radii are derived 
and the effects of mid-plane settling are discussed using the best-fitting accretion disk models. Finally the 
results of this paper are summarized in Section 6.

\section{Observations and Analysis}

\subsection{The Upper Scorpius Membership Sample}

The 12 Upper Scorpius infrared excess sources observed with SpeX were selected from the mid-infrared photometric 
and spectroscopic surveys of Carpenter et al. (2006) and Dahm \& Carpenter (2009), respectively. These sources
represent half of the late-type, disk-bearing stars identified by Carpenter et al. (2006) and include all K-type 
and all M0 through M2-type excess sources. To further place this sample into context with the low-mass stellar
population of Upper Scorpius, there are $\sim$250 known pre-main sequence stars in the mass range from $\sim$0.1 
to $\sim$2.0 M$_{\odot}$ (Preibisch \& Mamajek 2008; Preibisch et al. 2002). If the infrared excess fraction 
for late-type stars (19\%) derived by Carpenter et al. (2006) is assumed, $\sim$48 of these should host primordial 
disks. The SpeX sample therefore represents $\sim$25\% of all disk-bearing stars expected among the known
late-type association members. Integrating the best-fitting mass function for Upper Scorpius, Preibisch \& Mamajek (2008)
estimate that the total low-mass (0.1--2.0 M$_{\odot}$) stellar population of the association exceeds 2400 stars,
corresponding to a total disk-bearing population of $\sim$450 late-type stars. While the SpeX sample considered 
here does represent a substantial fraction of the late-type, disk-bearing stars identified by Carpenter et al. (2006),
it cannot be considered a statistically significant representation of the total low-mass stellar population of the 
association.

Given that unresolved companions could provide an explanation for transition-like SEDs (Ireland \& Kraus 2008),
some knowledge of the binary frequency of the Upper Scorpius late-type stellar population is warranted.
K\"{o}hler et al. (2000) used speckle interferometry and direct imaging for 118 X-ray selected T Tauri stars
in the greater Scorpius-Centaurus OB association to identify companions with separations ranging from 0.13\arcsec\ 
to 6.0\arcsec. They found a multiplicity fraction of $\sim$32.6\%, which exceeds the binary fraction of
main sequence stars by a factor $\sim$1.6, but is slightly lower than that observed in Taurus-Auriga
(K\"{o}hler et al. 2000). Kraus et al. (2008) found a similar frequency of binary companions among 82
late-type (G0--M4) Upper Scorpius members,  $\sim$35$\pm$5\%. Two of the disk candidates observed with SpeX
are established binaries, [PZ99]J161411.1-230536 and ScoPMS 31, resolved by high angular resolution imaging.
With projected separations of $\sim$32 and 84 AU, respectively (Metchev \& Hillenbrand 2009; K\"{o}hler et al. 2000),
it is unlikely that these stellar companions have significantly influenced the inner disk evolutionary
timescales of their primaries. Kraus et al. (2008) observed J160823.2-193001 and J160900.7-190852
using an aperture mask interferometry technique, placing firm upper limits that preclude the possibility
of an undetected stellar companion within $\sim$5 AU of these stars. The multiplicity fraction among the 
remaining Upper Scorpius disk-bearing sample is relatively unexplored, but is not expected to deviate
significantly from that derived by K\"{o}hler et al. (2000) and Kraus et al. (2008) for the pre-main sequence
population as a whole.

Shown in Figure 1 is the $J-H$, [8.0]$-$[4.5] color-color diagram for the 218 Upper Scorpius members included 
in the Carpenter et al. (2006) {\it Spitzer} IRAC and IRS survey. The 12 sources observed with SpeX span the 
full range of the 8.0 $\mu$m excess distribution, from moderate excess (e.g. J160643.8-190805, ScoPMS 31) 
to significant (e.g. J160900.7-190852). Their masses, derived using the pre-main sequence models of Siess et al. (2000) 
and assuming a distance of 145 pc, range from $\sim$0.25 to 2.0 M$_{\odot}$. General properties of the
stellar sample including spectral type, extinction ($A_{V}$), mass, radius, luminosity, and effective 
temperature are presented in Table 1.

\subsection{SpeX Observations}

The SpeX (Rayner et al. 2003) observations were made on the nights of 2009 May 20--23 under variable cirrus
and seeing conditions (0.6--0.9''). The Upper Scorpius members were observed in the short cross-dispersed SXD 
(0.84--2.4 $\mu$m) and long cross-dispersed LXD (2.2--5.2 $\mu$m) modes for complete near-infrared wavelength 
coverage. Bright A0 V stars were observed at similar airmasses ($\Delta$$\chi$ $<$ 0.1) and within short
periods of time of the program stars for telluric correction. Arc-lamps and internal flat-field exposures 
were obtained for each set of observations to account for instrument flexure. All observations were made 
with the 0.5\arcsec\ slit yielding a nominal spectral resolution of $\lambda$/$\delta$$\lambda$$\sim$1500. 

The spectra were reduced using {\it Spextool}, an IDL-based reduction package that provides for sky-subtraction, 
flat-fielding, wavelength calibration, and optimal extraction (Cushing et al. 2004). For telluric corrections 
{\it xtellcor}, an extension package of {\it Spextool}, was used to create a kernel for the telluric spectrum 
using a model spectrum of $\alpha$ Lyra. {\it xtellcor} interpolates over broad hydrogen absorption lines in 
the spectra of the A0 V stars using a technique developed by Vacca et al. (2003). The orders of the telluric
corrected spectra were then combined and the SXD and LXD spectra merged using routines available within 
{\it Spextool}. The resulting continuum levels across all orders and between the SXD and LXD modes were found 
to be remarkably consistent. The near infrared spectra were then merged with the {\it Spitzer} IRS spectra. 
Details of the {\it Spitzer} IRS observations and their subsequent reduction and analysis can be found in 
Dahm \& Carpenter (2009).

\section{Near-Infrared Continuum Excess Spectra}

The extinction-corrected, near-infrared (0.84--5.2 $\mu$m) spectra of the 12 Upper Scorpius sources are shown 
in Figure 2. Regions of strong telluric absorption (atmospheric transmission $<$20\%) have been excised from 
the figure. Also shown are the spectra of solar metallicity, main sequence stars of identical or closely matched
($\pm$1 sub-class) spectral type obtained from the IRTF spectral library (Rayner et al. 2009). These standards 
have reliable spectral types and are directly linked to the MK classification system. The standard or template 
spectra are scaled to the flux levels of the Upper Scorpius sources at 1.65 $\mu$m, near the peak of the
stellar SED and where extinction effects are minimized (D'Alessio et al. 1999; Furlan et al. 2006).

To characterize excess emission attributable to the inner disk rim, the scaled photospheric template spectra 
are subtracted from the dereddened spectra of the Upper Scorpius sample. The resulting continuum excess 
spectra are shown in Figure 3. Significant noise is present due to imperfect telluric correction, particularly 
in the thermal region where the water column varies both temporally and with airmass. In general, the slopes 
of the continuum excess spectra are increasing from $K-$band to $\sim$3.5 $\mu$m before turning over toward 
redder wavelengths. The shapes of many of the continuum excess spectra are suggestive of having been produced by 
single-temperature blackbodies. By fitting these excess spectra with Planck functions, the characteristic 
temperatures of the blackbodies are found to range from the sublimation temperature for silicate dust, i.e. 
$\sim$1400 K, to $\le$500 K. The slope of excess emission, particularly between $\sim$2.75 and 4.2 $\mu$m, was 
found to be of critical importance when fitting the blackbody profiles to the continuum excess spectra. 

To estimate the uncertainty associated with the derived blackbody temperatures, photospheric templates within 
a range of $\pm$1 sub-class of the assigned spectral type were subtracted from the Upper Scorpius spectra.
The best-fitting blackbody curves for this range of photosphere-subtracted excess spectra suggest typical 
uncertainties of $\sim$$\pm$200 K for most sources. The adopted blackbody curve as well as curves representing
the limits of uncertainty in the characteristic temperature are shown in Figure 3. The characteristic temperatures
of the best-fitting blackbodies are provided in Table 1 and are assumed to be representative of the dust 
temperatures in the disk interiors.

The disk fractions of Upper Scorpius and Taurus-Auriga differ significantly for the range in stellar masses 
considered here: 19\% for the former (Carpenter et al. 2006) and up to 75\% for the latter (Luhman et al. 2010). 
Muzerolle et al. (2003) used SpeX in LXD-mode to observe 9 CTTSs in Taurus-Auriga, 8 of which span a range 
of spectral types (G5--M1), luminosities (12.8--0.5 L$_{\odot}$), and stellar radii (3.6--1.8 R$_{\odot}$) that are 
comparable with those of the Upper Scorpius disk sample. Restricting further comparison of these samples to
just those sources having similar physical properties (i.e. spectral type, luminosity), the spectral profiles
of continuum excess emission of the Upper Scorpius disks are found to differ from those of Class II sources 
in Taurus-Auriga. For 6 of 9 members in the reduced Upper Scorpius sample, the characteristic temperatures 
of excess emission fall below the dust sublimation temperature, a condition satisfied by
only one member (the binary DQ Tau) of the Taurus-Auriga sample. Furthermore, the excess profiles of 
J160643.8-190805 and J160357.9-194210, which are not represented in Figure 3, exhibit no distinctive shape, 
implying minimal continuum excess emission at near-infrared wavelengths. Shallow rises may be present near 
$\sim$3.75 $\mu$m for these 2 sources, suggestive of warm ($\le$500 K) dust emission, but higher SN spectra 
are needed for confirmation. The excess spectra of J161420.2-190648, [PZ99]J160357.6-203105, J160900.7-190852, 
and J155829.8-231007 all have characteristic temperatures at or near the dust sublimation temperature, 
implying that emission arises from dust in close proximity to the host star. Of these sources, all are 
suspected accretors (Section 4). 

\section{Accretion Luminosity and Mass Accretion Rates}

Dahm \& Carpenter (2009) identified 5 Upper Scorpius members in the present sample as probable accretors using 
the H$\alpha$ velocity width criteria of White \& Basri (2003): [PZ99]J160357.6-203105 (K5), J160900.7-190852 (K7),
J161420.2-190648 (M0), ScoPMS 31 (M0.5), and J155829.8-231007 (M3). The near-infrared spectra of these sources
exhibit \ion{He}{1} $\lambda$10830, Pa$\gamma$, and Pa$\beta$ emission and for most, weak Br$\gamma$ emission. 
Shown in Figure 4 are the normalized spectra of these sources centered upon Pa$\gamma$, Pa$\beta$, and 
Br$\gamma$. Dahm \& Carpenter (2009) used veiling at $\lambda$6500 \AA\ as well as \ion{Ca}{2} $\lambda$8542 
emission line luminosity to estimate mass accretion rates ($\dot{M}$) for these stars. In the near-infrared,
Muzerolle et al. (1998) found Pa$\beta$ and Br$\gamma$ emission line luminosity ($L_{Pa\beta}$ and $L_{Br\gamma}$) 
to be well-correlated with accretion luminosity, $L_{acc}$. The resulting least-squares fits from Muzerolle et al. (1998)
are given by:

log$(\frac{L_{acc}}{L_{\odot}})=(1.14\pm0.16)$log$(\frac{L_{Pa\beta}}{L_{\odot}})+3.15\pm0.58$   (1)

\noindent and

log$(\frac{L_{acc}}{L_{\odot}})=(1.26\pm0.19)$log$(\frac{L_{Br\gamma}}{L_{\odot}})+4.43\pm0.79$   (2).

For the 5 suspected accretors in the Upper Scorpius sample, template spectra of identical or similar spectral
type from the IRTF spectral library of Rayner et al. (2009) were used to subtract Pa$\beta$ and Br$\gamma$ 
photospheric absorption. Pa$\beta$ and Br$\gamma$ line luminosities were then determined using their measured 
equivalent widths and the extinction-corrected $J-$ and $K_{S}-$band magnitudes obtained from the 2MASS point 
source catalog. The resulting emission line luminosities were then transformed into $L_{acc}$ using the above 
linear relationships of Muzerolle et al. (1998). The derived $L_{acc}$ values are directly proportional to 
$\dot{M}$ such that:

$L_{acc}\sim \frac{GM_{*}\dot{M}}{R_{*}}(1-\frac{R_{*}}{R_{in}})$   (3)

\noindent where the stellar mass and radius estimates used are those predicted by the pre-main sequence models
of Siess et al. (2000). The factor of ($1 - R_{*}/R_{in}$) is assigned a value of 0.8, which assumes an inner 
disk radius ($R_{in}$) of 5 R$_{*}$ (Gullbring et al. 1998). Given the possibility that inner disk radii for 
the Upper Scorpius sources may exceed those of typical Class II sources in Taurus-Auriga, this value could
be underestimated by a factor of $\sim$1.25. The derived $L_{acc}$ and $\dot{M}$ values with their associated 
uncertainties are presented in Table 2. These uncertainties arise from multiple sources: error in the measured 
equivalent width (assumed to be $\sim$20\%), spectral type uncertainty when correcting for photospheric 
absorption ($\pm$1 sub-class), the uncertainty in each coefficient of the Muzerolle et al. (1998) relationships,
and the uncertainty in R$_{in}$ when determining $\dot{M}$ values. 

In general the $\dot{M}$ values derived using $L_{Pa\beta}$ and $L_{Br\gamma}$ agree reasonably well with each 
other and with the $\dot{M}$ values from the veiling and \ion{Ca}{2} $\lambda$8542 analysis of Dahm \& Carpenter (2009). 
In summary, 5 of 12 of the Upper Scorpius sample are accreting, providing unambiguous evidence for the presence of 
gas within the terrestrial regions of these disk-bearing systems.

\section{Accretion Disk Model Analysis}

To further examine inner disk structure and to constrain inner disk radii for the Upper Scorpius 
disk-bearing sample, the observed SEDs are compared with those predicted by the accretion disk models 
of Robitaille et al. (2006). To facilitate the comparison, the flux-calibrated SpeX observations 
presented here are merged with the {\it Spitzer} IRS low-resolution spectra and MIPS 24 and 70 $\mu$m fluxes taken 
from Dahm \& Carpenter (2009) and Carpenter et al. (2009), respectively. The resulting 0.8--70 $\mu$m
SEDs are well sampled between 2.2 and 24 $\mu$m, a spectral region dominated by disk emission originating 
from the terrestrial region. 

The Robitaille et al. (2006) grid of pre-computed, two-dimensional radiative transfer models consists of 
20,000 young stellar objects in varying stages of evolution and viewed from 10 inclination angles. A total 
of 14 parameters are randomly sampled that specify stellar (e.g. $M_{*}$, $R_{*}$, $T_{eff}$) as well as 
disk (e.g. $M_{disk}$, $\dot{M}$, $R_{min}$) properties. The models have been successfully applied to the 
SEDs of 30 spatially resolved Class I and II sources in Taurus-Auriga by Robitaille et al. (2007), including 
the transition disk objects GM Aur and DM Tau.

The SEDs of 9 Upper Scorpius members were compared to the accretion disk models. The sources excluded from 
the model fitting analysis were J160545.4-202308 and J155829.8-231007, which were not observed with IRS, 
and [PZ99]J160421.7-213028, which exhibited significant (factor of $\sim$4) mid-infrared variability
(Dahm \& Carpenter 2009). To compare the observed SEDs with the accretion disk models, fluxes were measured
in $\sim$30 narrow passbands defined between 0.8 and 37.0 $\mu$m. The MIPS 24 and 70 $\mu$m fluxes were 
also incorporated, with the 70 $\mu$m fluxes being used to constrain disk emission originating from beyond 
the terrestrial region. 

Two assumptions were made in the model fitting analysis: that all Upper Scorpius members lie between 125 
and 185 pc distant, and that foreground extinction is no more than $A_V \sim 3$ mag, both reasonable 
assumptions based upon substantial Upper Scorpius literature (e.g. Preibisch \& Zinnecker 1999; 
Preibisch et al. 2002; Carpenter et al. 2006). A range of effective temperatures was defined for each 
Upper Scorpius source based upon its adopted spectral type and assuming an uncertainty of $\pm$1 
spectral sub-class. Only models having effective temperatures within this specified range were considered 
when fitting the observed SEDs. The model fitting program was written in IDL and uses the $\chi^{2}$ statistic 
to evaluate the goodness of fit of each model for each assumed distance, extinction value, and inclination 
angle. Table 3 lists the range of $T_{eff}$ values considered for each Upper Scorpius source, the number of 
models included within the specified range, the best-fitting model number, its corresponding $A_{V}$, distance,
and inclination angle, and the minimum reduced $\chi^{2}$ value achieved by the fit. 

Shown in the upper panels of Figures 5a--i are the observed 0.8--70.0 $\mu$m SEDs for the 9 Upper Scorpius 
sources included in the model fitting analysis. Superimposed in red are the best-fitting model profiles 
placed at their respective distances and extinctions. The disk contributions to the model SEDs are represented 
by the dotted black curves, and the model stellar atmospheres, taken from the Kurucz (1979) atlas, are
plotted in blue. In general, reasonable fits were achieved for many of the Upper Scorpius disk-bearing
stars despite the modest number of models (typically $\le$2000) examined for each source. The near-infrared
spectrum of the strongly accreting source J161420.2-190648 is suggestive of significant extinction, 
$A_V\sim$6 mag, (Figure 5f). The optimal model fits for this source yielded the largest $\chi^{2}$ values 
of the sample given the a priori assumption that $A_V \le 3$ mag.

Degeneracy among model SEDs having significantly different stellar and disk properties is an inherent 
limitation of the SED fitting process. It is possible, however, to place constraints upon a specific 
parameter by examining the range of values returned by the best-fitting models. To identify this subset
of models, $\chi^{2} - \chi^{2}_{min} < N$ is used as a statistical measure of goodness of fit, where 
$\chi^{2}_{min}$ is the minimum $\chi^{2}$ value returned by the fitting process and $N$ is set to the 
1-$\sigma$ confidence interval for $\nu$ degrees of freedom. In Table 4 the disk properties 
of the Upper Scorpius sample obtained from these best-fitting subsets are summarized. Tabulated for each source are 
the minimum, best-fitting, and maximum values of disk mass (M$_{disk}$), $\dot{M}$, and inner disk radius 
(R$_{in}$). The ranges of M$_{disk}$ and $\dot{M}$ are significant, up to several orders of magnitude
separating the minimum and maximum values for a given source. The observed dispersion in M$_{disk}$ arises 
in part from the lack of far-infrared or millimeter wavelength observations that could constrain outer 
disk emission where substantial quantities of disk mass could remain unaccounted for. Mass accretion rates 
calculated by the models were found by Robitaille et al. (2007) to be overestimated in their sample of 
Taurus sources. In the Upper Scorpius sample, whether established accretors or not, $\dot{M}$ is 
observed to range at least 3 orders of magnitude, and in some cases up to 5. For most suspected accretors 
included in the model fitting analysis, the $\dot{M}$ values derived in Section 4 using the Pa$\beta$ and 
Br$\gamma$ emission line luminosities are consistent with those predicted by the best-fitting models.

As might be expected, inner disk radius appears to be better constrained by the subsets of best-fitting 
models than either M$_{disk}$ or $\dot{M}$. Several Upper Scorpius sources, e.g. [PZ99]J161411.0-230536, 
[PZ99]J160357.6-203105, and ScoPMS 31, exhibit ranges in $R_{in}$ from minimum to maximum of an order of 
magnitude or less. Five sources have best-fitting inner disk radii that exceed their respective dust 
sublimation radii: [PZ99]J161411.0-230536 ($R_{in}$=0.87 AU), J160900.7-190852 (2.18 AU), ScoPMS 31 (9.62 AU), 
J161115.3-175721 (0.29 AU), and J160357.9-194210 (0.17 AU). Given the poor quality of the $\chi^{2}$ fit
for the strongly accreting source J161420.2-190648, the best-fitting $R_{in}$ value of 0.8 AU is regarded 
with skepticism. The remaining Upper Scorpius sources ([PZ99]J160357.6-203105, J160643.8-190805, and 
J160823.2-193001) have best-fitting inner disk radii that are consistent with their sublimation radii.
The confidence levels for these inner disk radii, however, are best examined in terms of a probability 
density function determined from the $\chi^{2}$ values of the accretion disk model fits. 

\subsection{Constraints for Inner Disk Radii}

Flux from the terrestrial disk regions is well sampled by the SpeX and IRS observations, which
are dominated by disk emission extending from the sublimation radius to more than 20 AU for a typical 
$\sim$0.7 M$_{\odot}$ pre-main sequence star. The Upper Scorpius disks are presumably at an advanced 
evolutionary stage relative to those found around Class II sources in Taurus-Auriga and exhibit SEDs 
that are consistent with reduced levels of near and mid-infrared disk emission (Dahm \& Carpenter 2009). 
Robitaille et al. (2007) find that all Taurus sources, except for the known transition disk objects 
(e.g GM Aur, DM Tau) can be fit by models having disks and envelopes with inner disk radii equal to 
the dust sublimation limit. Approximately one-third of the Robitaille et al. (2006) models have inner 
disk radii set to the dust destruction radius. The remaining models have increasing inner disk radii 
that span from the dust destruction radius to 100 AU. The inner disk gaps are treated by the models as 
being completely evacuated of dust (Robitaille et al. 2006).

To constrain R$_{in}$ for the Upper Scorpius disk sample a probability ($P$) is calculated for a given
inner disk radius using the returned $\chi^{2}$ values from the model fits of the observed SEDs:

$P=\frac{1}{\sqrt{2 \pi}} e^{\frac{-\chi^{2}}{2}}$   (4)

\noindent This probability density function assumes a normal distribution for the returned $\chi^{2}$ 
values of the individual fits. In the center panels of Figures 5a--i, $P$ is plotted as a function of 
R$_{in}$ for each Upper Scorpius source. Superimposed in the figures as cross-hatched histograms are 
the distributions of inner disk radii for the entire sample of models considered for each source. 
Only 3 Upper Scorpius disks have predicted inner radii that are larger than their respective dust 
sublimation radii at the 1$\sigma$ confidence level or greater: [PZ99]J161411.0-230536 (2-$\sigma$), 
J160900.7-190852 (1-$\sigma$), and ScoPMS31 (3-$\sigma$). The remaining sources have inner disk radii 
that are either most consistent with the sublimation radius or exhibit a broad dispersion among predicted 
$R_{in}$ values (e.g. J161115.3-175721). The SED of J160643.8-190805,
which exhibits minimal near-infrared continuum excess emission (Section 3), is readily fit by a large number 
of models having a range of inner disk radii. This source exhibits an SED that is reminiscent of those 
defined as anemic (Lada et al. 2006), weak (Dahm \& Hillenbrand 2007), or homologously depleted
disk-bearing systems (Currie et al. (2009).

\subsection{Disk Mid-plane Settling}

The current disk evolution scenario suggests that dust grains coagulate and settle toward the mid-plane
prior to the formation of large planetesimals. The Robitaille et al. (2006) models use 2 disk structure 
parameters to mimic the effects of dust settling: a disk flaring parameter ($\beta$) and a disk scale 
height factor ($z$). If both of these parameters are low for a given model, this could be indicative of 
dust mid-plane settling (Robitaille et al. 2006). In the lower panels of Figures 5a--i $\beta$ is plotted
as a function of $z$ for the entire sample of models considered for each source. Superimposed in red are 
the best-fitting models as determined by the $\chi^{2} - \chi^{2}_{min} < N$ measure of goodness of fit. 
In general the spread in both parameters for the best-fitting models is significant, suggesting that 
neither is well-constrained. Some argument can be made that the SEDs of [PZ99]J161411.0-230536,
J160643.8-190805, and ScoPMS31 are better fit by models having lower than average values of $\beta$. 
The disk scale height factors for these sources, however, are found to vary significantly. Evidence for
dust settling effects among the Upper Scorpius primordial disk sample remains inconclusive at best.

\section{Discussion and Conclusions}

The clearly different near- and mid-infrared color excess distributions of the Upper Scorpius and
Taurus-Auriga Class II populations found by Dahm \& Carpenter (2009) suggest that the Upper Scorpius 
stars have experienced some degree of inner disk evolution relative to their presumably younger 
counterparts in Taurus-Auriga. This is circumstantially supported by the characteristic temperatures 
of the continuum excess emission for most Upper Scorpius sources, which are substantially lower than 
the sublimation temperature for silicate dust. This condition is satisfied by only one member of the 
Muzerolle et al. (2003) Taurus-Auriga sample of Class II sources having similar spectral types and 
luminosities. The accretion disk model fitting results, however, are unable to effectively constrain 
inner disk radii for the majority of the Upper Scorpius disk-bearing sample. This could in part result 
from the limited number of models examined for each source, which do not sample the full range of 
available parameter space of disk structure. Another limitation of the Robitaille et al. (2006) models 
is the assumption that disk gaps are completely devoid of dust, a reduction in complexity that impacts 
the predicted near- and mid-infrared excess distributions, critical for constraining disk emission 
originating from the inner disk rim.

In summary 3 of 9 late-type, disk-bearing stars in the Upper Scorpius sample exhibit SEDs that are most 
consistent with having inner disk radii that lie beyond the sublimation radius for silicate dust: 
[PZ99]J161411.0-230536, J160900.7-190852, and ScoPMS31. The best-fitting models for 2 additional
sources: J161115.3-175721 and J160357.9-194210, have inner disk radii that substantially exceed their 
respective dust sublimation radii, but the probability distributions of $R_{in}$ values for these 
sources exhibit significant dispersion. Proposed indicators for the effects of mid-plane settling
in the best-fitting Robitaille et al. (2007) models of the Upper Scorpius sample (i.e. concurrent 
decreased values of the disk flaring parameter and disk scale height factor) are inconclusive, leaving 
open to question the nature of the remaining Upper Scorpius primordial disks and their evolutionary state. 

Transition disks are believed to represent an early stage of disk clearing and may be in the process 
of rapid grain growth, planetesimal and planet formation. Adopting the disk classification scheme of
Luhman et al. (2010), many Upper Scorpius sources would be classified as pre-transitional objects, 
i.e. disks that exhibit reduced emission at wavelengths $\le$10 $\mu$m, but that still retain 
significant disk emission at longer wavelengths. Two of the Upper Scorpius pre-transitional candidates 
appear to exhibit gapped disk structure: J160900.7-190852 and ScoPMS31. Both sources are accreting 
(Dahm \& Carpenter 2009 and Section 4) and both exhibit SEDs that are most consistent with having 
substantial inner disk radii, $\sim$1--10 AU. At least one Upper Scorpius source, J160643.8-190805,
exhibits an infrared SED that is suggestive of an homologously depleted disk system as defined by 
Currie et al. (2009).

Before attributing the reduced levels of infrared excess emission in the Upper Scorpius sample to 
disk evolutionary processes, binarity must be considered as an explanation for the cleared out inner 
cavities (e.g. CoKu Tau 4, Ireland \& Kraus 2008). Two of the disk candidates included in this analysis, 
[PZ99]J161411.1-230536 and ScoPMS 31, are established binaries resolved by high angular resolution 
imaging. It must also be acknowledged that the SpeX sample represents only $\sim$25\% of all primordial
disk-bearing stars expected among the $\sim$250 known late-type members of the Upper Scorpius OB association.
Observations of more late-type, disk-bearing systems are critically needed to confirm the results 
presented here. 

Improved modeling of the Upper Scorpius primordial disk-bearing sample is clearly needed to provide 
better constraints for the inner disk structure of these presumably evolved primordial disk systems. 
Monte Carlo three-dimensional, radiative transfer codes are now available that could be applied to 
the SEDs of these sources. High angular resolution imaging is also needed to identify close ($\le$10 AU) 
binary companions that may account for the reduced near- and mid-infrared excess emission of these 
sources relative to Class II sources in Taurus-Auriga. Infrared interferometric observations or precision 
radial velocity monitoring are also needed to identify tighter pairs or spectroscopic binaries that 
would be capable of dynamically clearing the disk interiors of these systems.

\acknowledgments
This work is based on observations made with the {\it Spitzer} Space Telescope, which is operated by the 
Jet Propulsion Laboratory (JPL), California Institute of Technology, under NASA contract 1407. The
Digitized Sky Surveys, which were produced at the Space Telescope Science Institute under 
U.S. Government grant NAG W-2166, were used as were the the SIMBAD database operated at CDS, 
Strasbourg, France, and the Two Micron All Sky Survey (2MASS), a joint project of the University of 
Massachusetts and the Infrared Processing and Analysis Center (IPAC)/California Institute of Technology, 
funded by NASA and the National Science Foundation. The author gratefully acknowledges John Carpenter and
an anonymous referee for many helpful suggestions that significantly improved this manuscript. The 
author also thanks Elise Furlan, Thayne Currie, and Joan Najita for insightful discussions and 
Thomas Robitaille for graciously providing direct access to the accretion disk models. Finally
the author thanks John Rayner for obtaining a second-epoch SpeX SXD spectrum of [PZ99]J160421.7-213028.

\clearpage
\begin{figure}
\epsscale{0.65}
\hspace{2cm}  \vspace{2cm}  \includegraphics[width=11cm,angle=0]{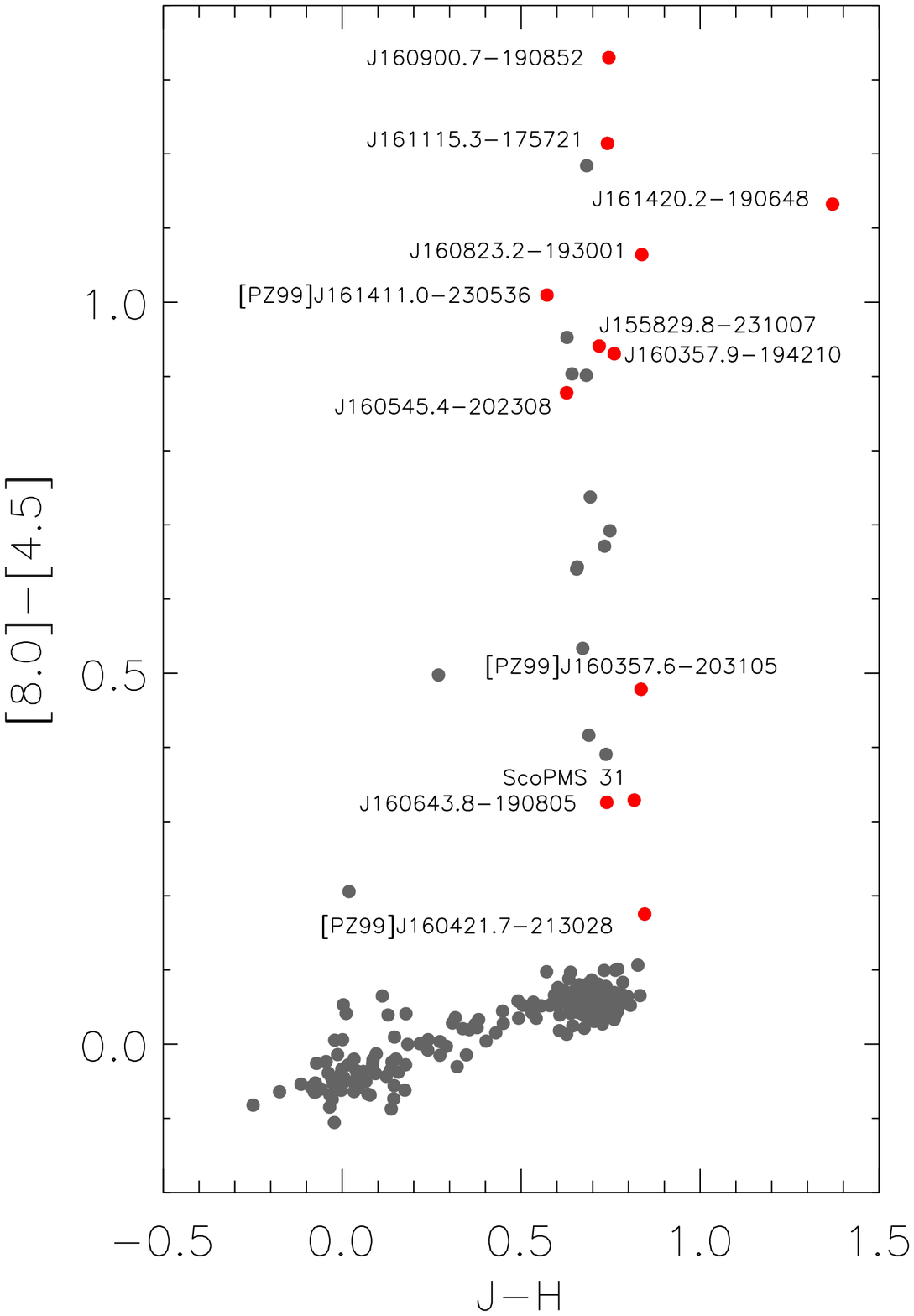}  \vspace{-2cm}    
\caption[f1.ps]{The $J-H$, [8.0]$-$[4.5] color-color diagram for 218 Upper Scorpius members from the {\it Spitzer} 
IRAC and IRS survey of Carpenter et al. (2006). Shown in red are the 12 late-type disk-bearing stars observed with 
SpeX. The sources span the full range of the 8.0 $\mu$m excess distribution, from slight excess (e.g. J160643.8-190805, 
ScoPMS 31) to significant (e.g. J160900.7-190852).
\label{f1}}
\end{figure}
\clearpage

\clearpage
\renewcommand{\thefigure}{\arabic{figure}\alph{subfigure}}
\setcounter{subfigure}{1}
\begin{figure}
\epsscale{0.5}
\hspace{2cm}  \vspace{2cm}  \includegraphics[width=11cm,angle=0]{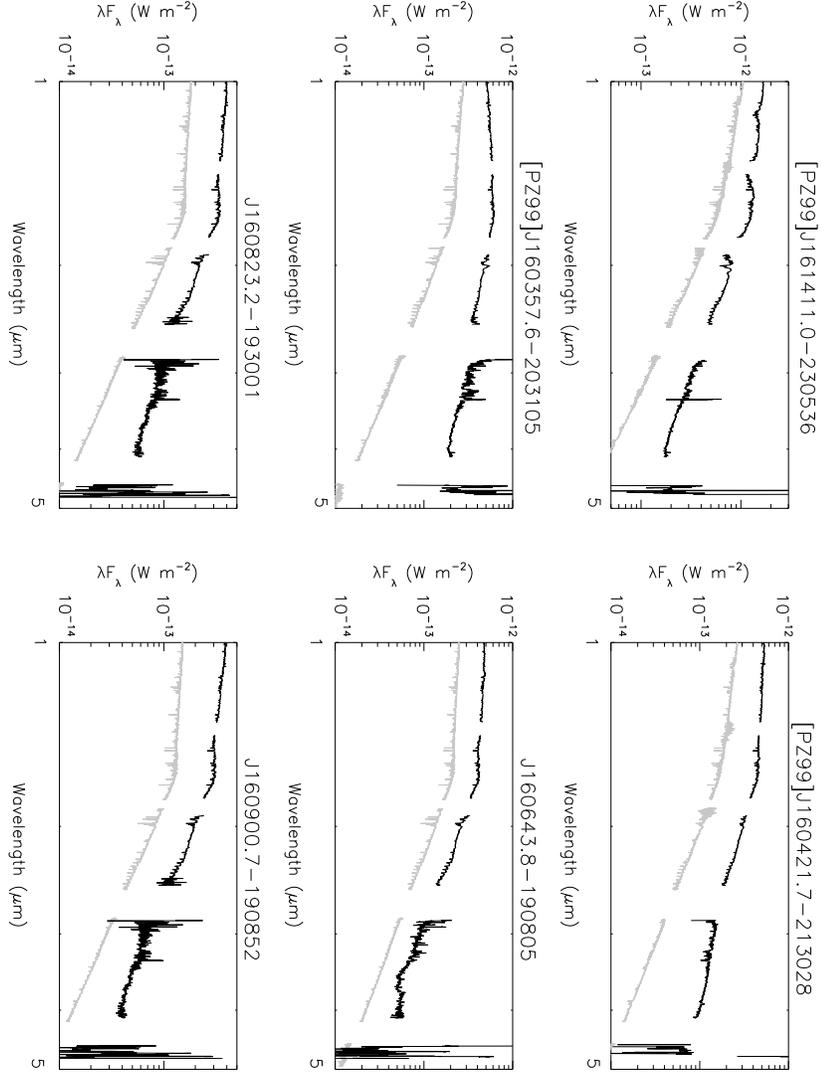} \vspace{-2cm}
\caption[f2a.ps]{The SpeX near infrared (0.84--5.2 $\mu$m) spectra of the 12 late-type disk-bearing
stars in Upper Scorpius, de-reddened by their individual extinctions tabulated in Dahm \& Carpenter (2009).
Superimposed in gray are the spectra of solar metalicity, main sequence stars of identical spectral type
or closely matched obtained from the IRTF spectral library (Rayner et al. 2009). The standard spectra are 
scaled to the flux levels of the Upper Scorpius sources near 1.65 $\mu$m, where stellar photospheric flux 
peaks and where extinction effects are minimized. For the purpose of display, the standard spectra are offset
from the object spectrai by an additive constant.
\label{f2a}}
\end{figure}
\clearpage

\addtocounter{figure}{-1}
\addtocounter{subfigure}{1}
\clearpage
\begin{figure}
\epsscale{0.5}
\hspace{2cm}  \vspace{2cm}  \includegraphics[width=11cm,angle=0]{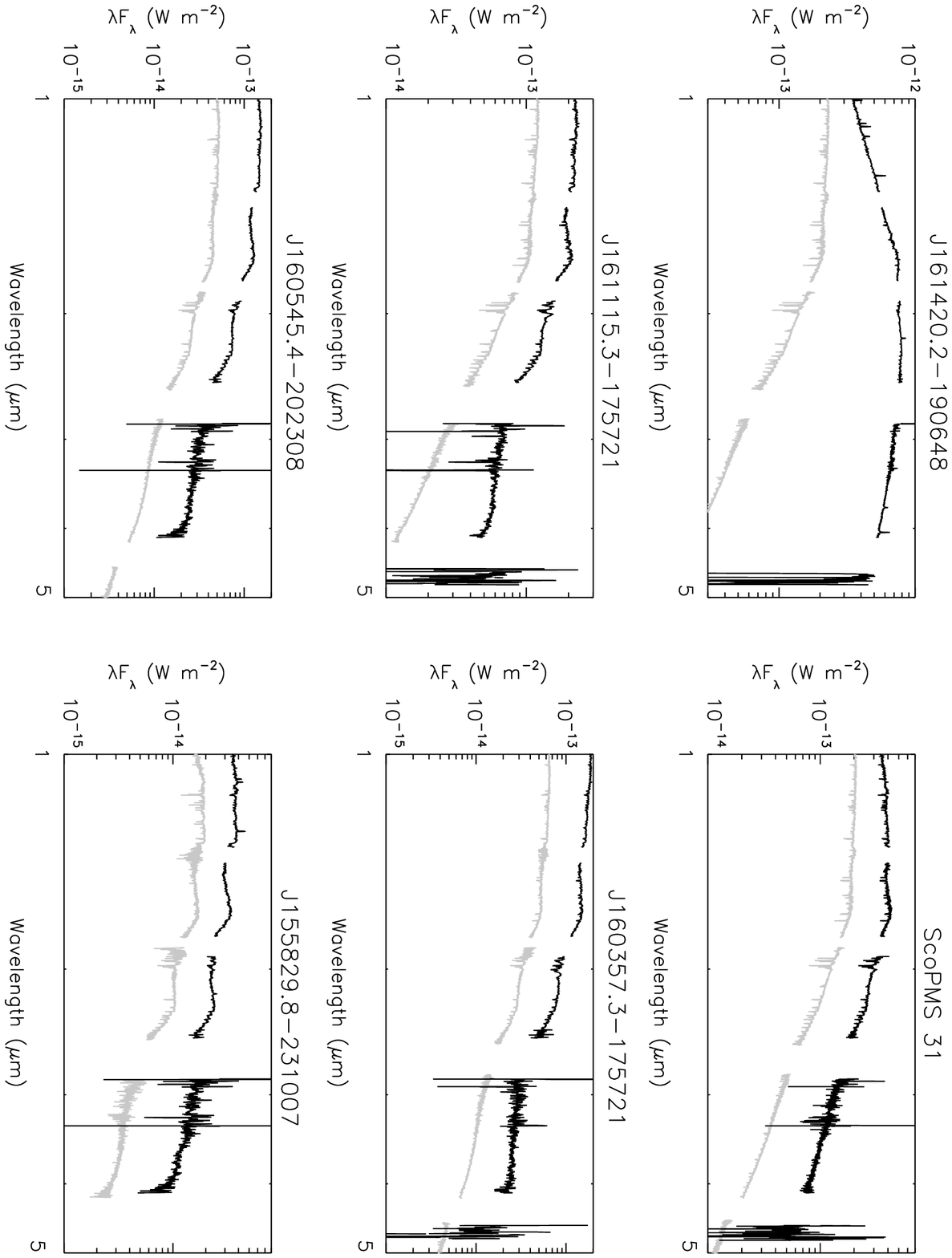}  \vspace{-2cm}    
\caption[f2b.ps]{(continued)
\label{f2b}}
\end{figure}
\clearpage

\clearpage
\renewcommand{\thefigure}{\arabic{figure}\alph{subfigure}}
\setcounter{subfigure}{1}
\begin{figure}
\epsscale{0.5}
\hspace{2cm}  \vspace{2cm}  \includegraphics[width=11cm,angle=0]{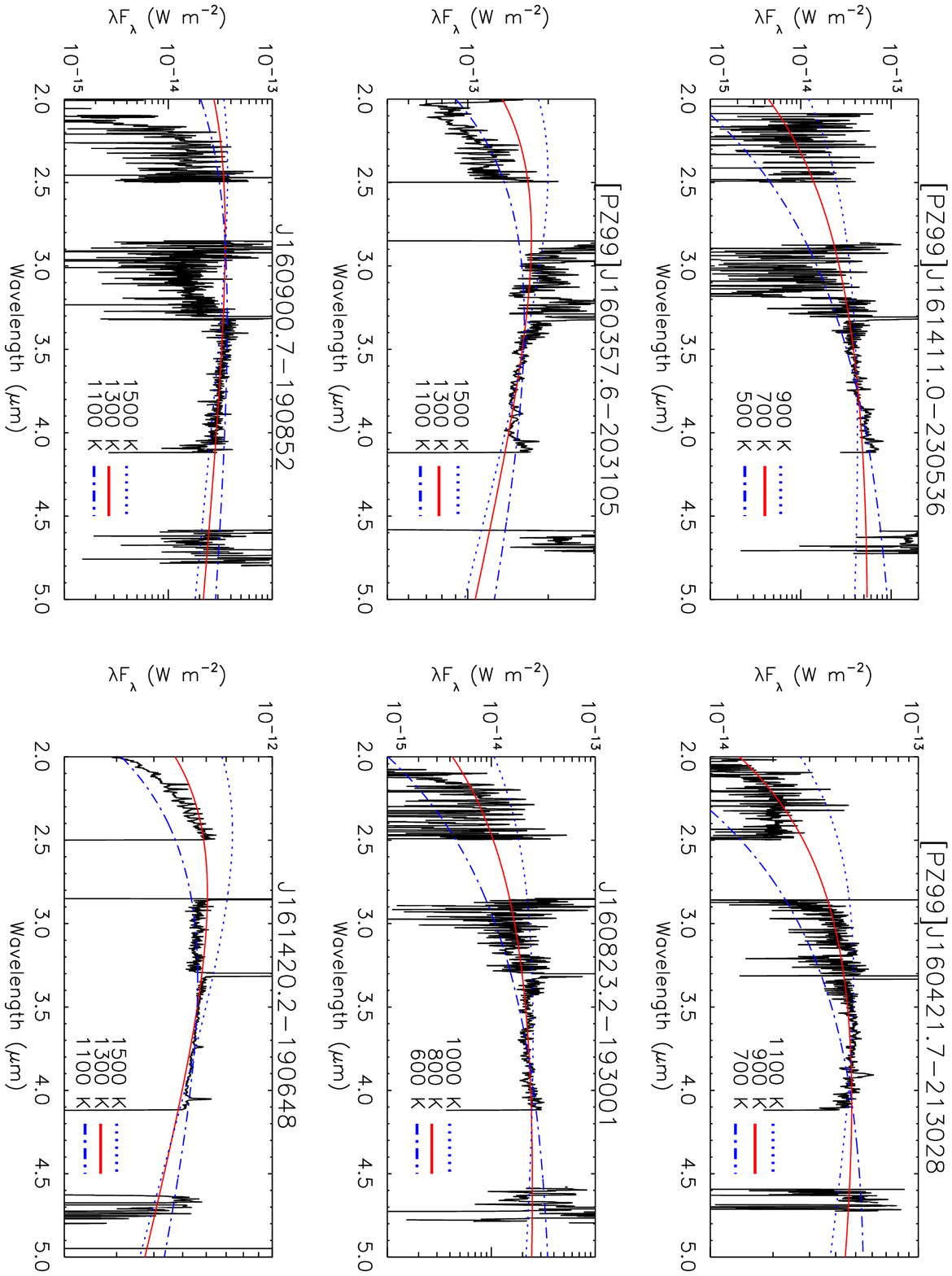} \vspace{-2cm}
\caption[f3a.ps]{The near-infrared continuum excess spectra for 10 Upper Scorpius late-type, disk-bearing
stars created by subtracting scaled photospheric template spectra obtained from the IRTF spectral library
(Rayner et al. 2009) from the dereddened object spectra. The template spectra are of identical spectral
type of the Upper Scorpius sources or closely matched, within $\pm$1 subclass. Significant
noise is present due to imperfect telluric correction, particularly in the thermal region where the water
column varies both temporally and with airmass. Shown in red are the Planck functions that best fit the
continuum excess spectra. Most have characteristic temperatures ranging from the dust sublimation 
temperature, near $\sim$1400 K, to less than $\sim$500 K. Also depicted (in blue) are blackbody curves 
representing the upper and lower limits of uncertainty associated with the blackbody characteristic 
temperatures.
\label{f3a}}  
\end{figure}
\clearpage

\addtocounter{figure}{-1}
\addtocounter{subfigure}{1}
\clearpage
\begin{figure}
\epsscale{0.5}
\hspace{2cm}  \vspace{2cm}  \includegraphics[width=11cm,angle=0]{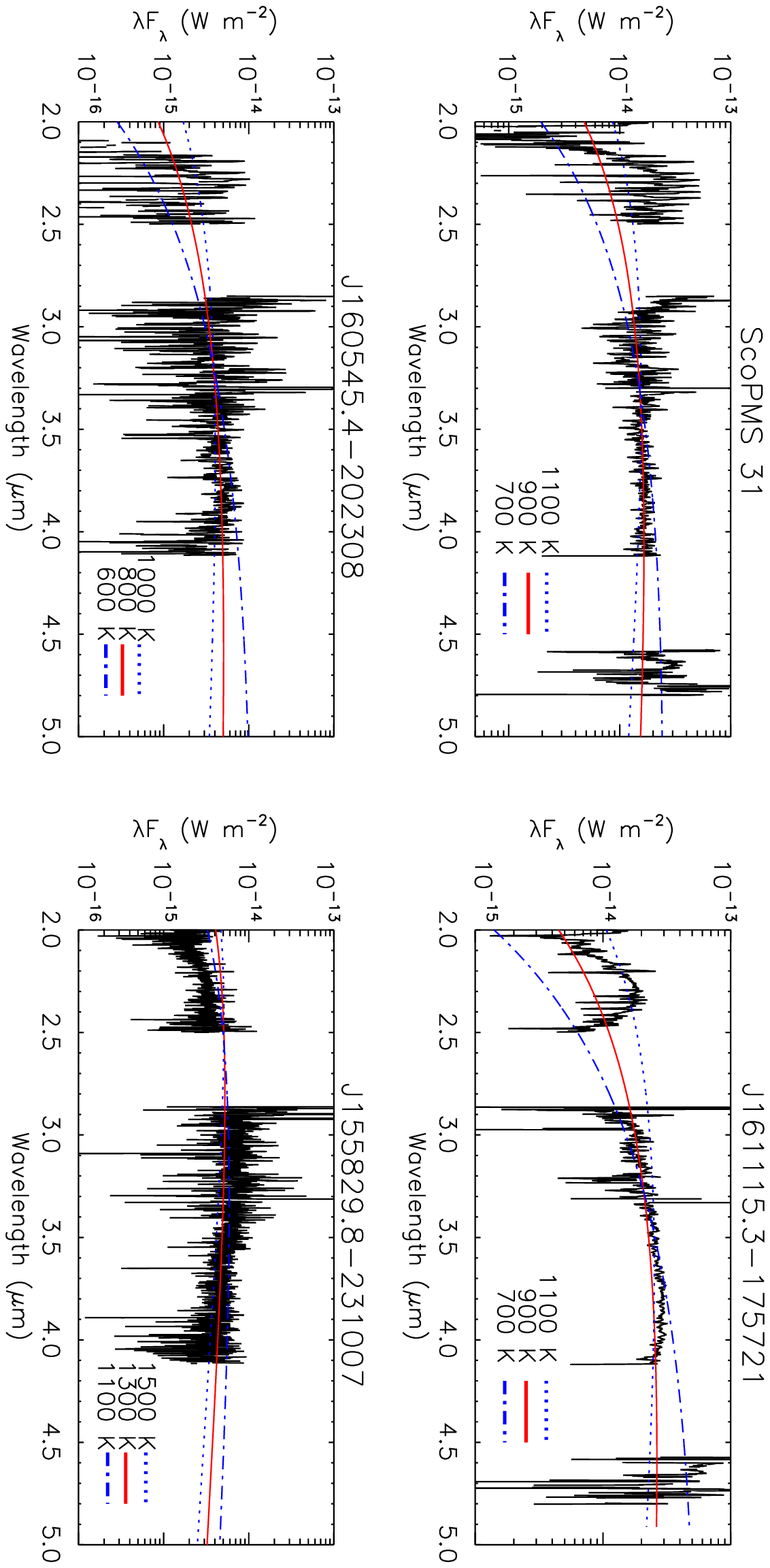}  \vspace{-2cm}    
\caption[f3b.ps]{(continued)
\label{f3b}}
\end{figure}
\clearpage

\clearpage
\begin{figure}
\epsscale{0.5}
\hspace{2cm}  \vspace{2cm}  \includegraphics[width=11cm,angle=0]{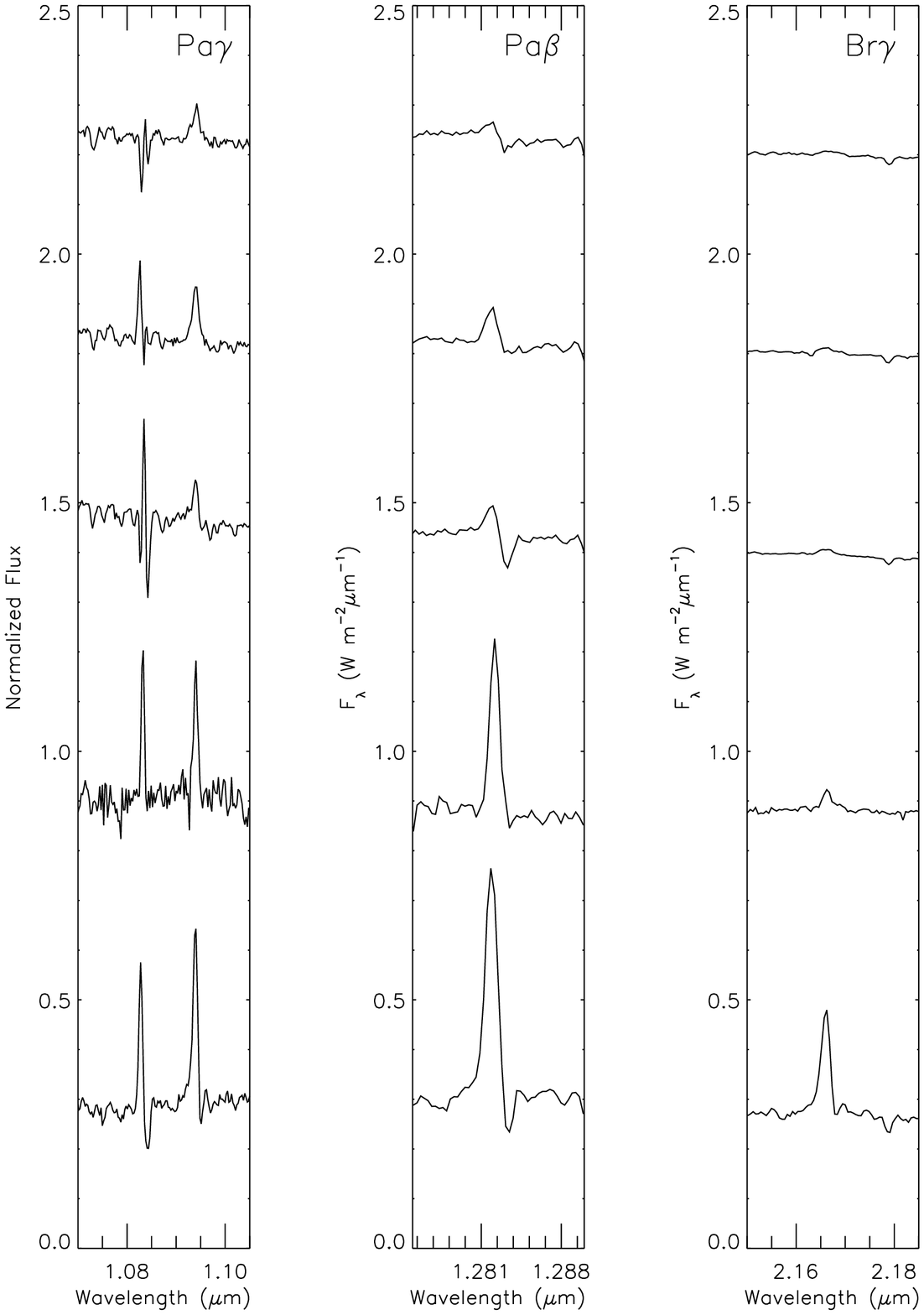}  \vspace{-2cm}    
\caption[f4.ps]{Emission features of \ion{He}{1} $\lambda$10830 and Pa$\gamma$ (left), Pa$\beta$ (center),
and Br$\gamma$ (right), for the 5 accreting sources in the sample. From top to bottom, the sources
shown are: ScoPMS 31 (M0.5), J160900.7-190852 (K7), [PZ99]J160357.6-203105 (K5), J155829.8-231007 (M3),
and J161420.2-190648 (M0).
\label{f4}}
\end{figure}
\clearpage

\clearpage
\renewcommand{\thefigure}{\arabic{figure}\alph{subfigure}}
\setcounter{subfigure}{1}
\begin{figure}
\epsscale{0.5}
\hspace{2cm}  \vspace{2cm}  \includegraphics[width=11cm,angle=0]{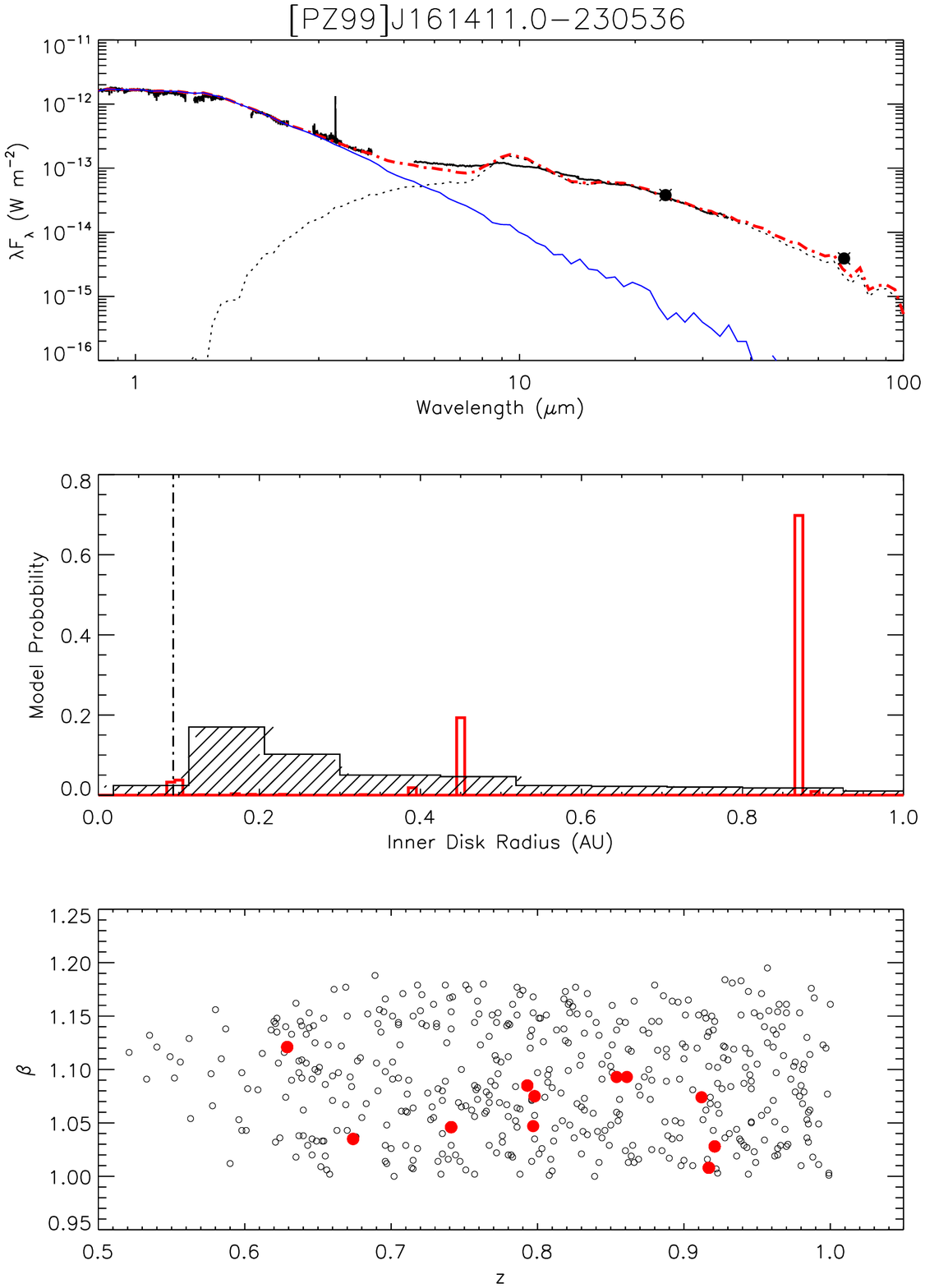} \vspace{-2cm}
\caption[f5a.ps]{(top) The observed 0.8--70.0 $\mu$m SEDs for 9 late-type Upper Scorpius disk-bearing
stars constructed from the SpeX 0.8--5.2 $\mu$m spectra, the {\it Spitzer} IRS 5.4--37.0 $\mu$m spectra,
and the MIPS 24 and 70 $\mu$m photometry. Superimposed in red are the best-fitting model profiles of
Robitaille et al. (2006), placed at their respective distances and extinctions. The disk contributions
to the model SEDs are shown as dotted black curves, and the model stellar photospheres from the 
Kurucz (1979) atlas as solid blue curves. (center) The probability ($P$) of inner disk radii ($R_{in}$) 
for the Upper Scorpius disk-bearing sample plotted as a function $R_{in}$. Superimposed in the figures 
as cross-hatched histograms are the distributions of inner disk radii for the entire sample of models 
considered for each source. The vertical dashed lines represent the dust sublimation radii for the 
best-fitting models. (bottom) The disk flaring parameter ($\beta$) plotted as a function of disk scale 
height factor ($z$) for all models examined in the SED fitting process. The best-fitting models are shown in red. 
\label{f5a}}
\end{figure}
\clearpage

\addtocounter{figure}{-1}
\addtocounter{subfigure}{1}
\clearpage
\begin{figure}
\epsscale{0.5}
\hspace{2cm}  \vspace{2cm}  \includegraphics[width=11cm,angle=0]{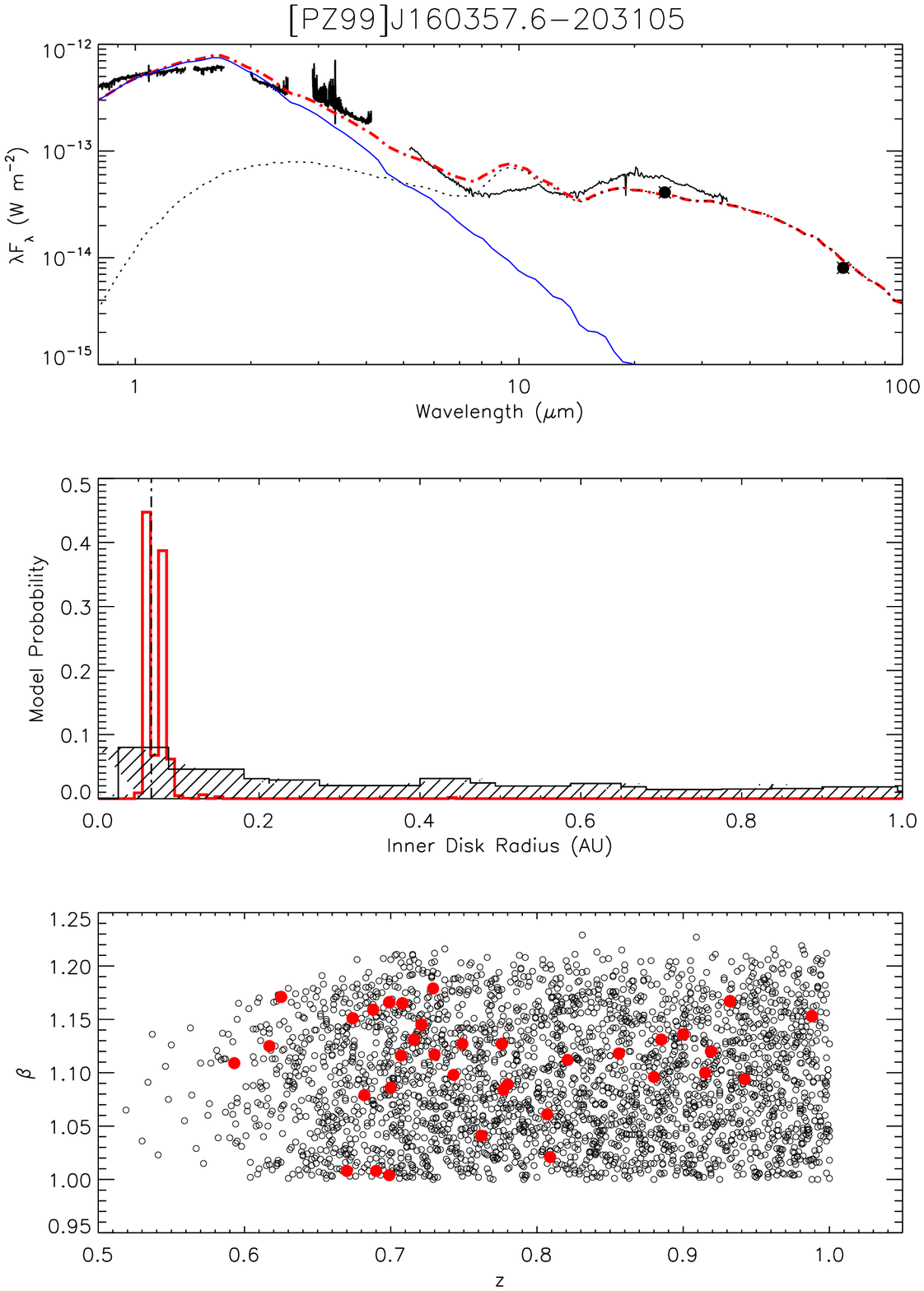}  \vspace{-2cm}    
\caption[f5b.ps]{(continued)
\label{f5b}}
\end{figure}
\clearpage

\addtocounter{figure}{-1}
\addtocounter{subfigure}{1}
\clearpage
\begin{figure}
\epsscale{0.5}
\hspace{2cm}  \vspace{2cm}  \includegraphics[width=11cm,angle=0]{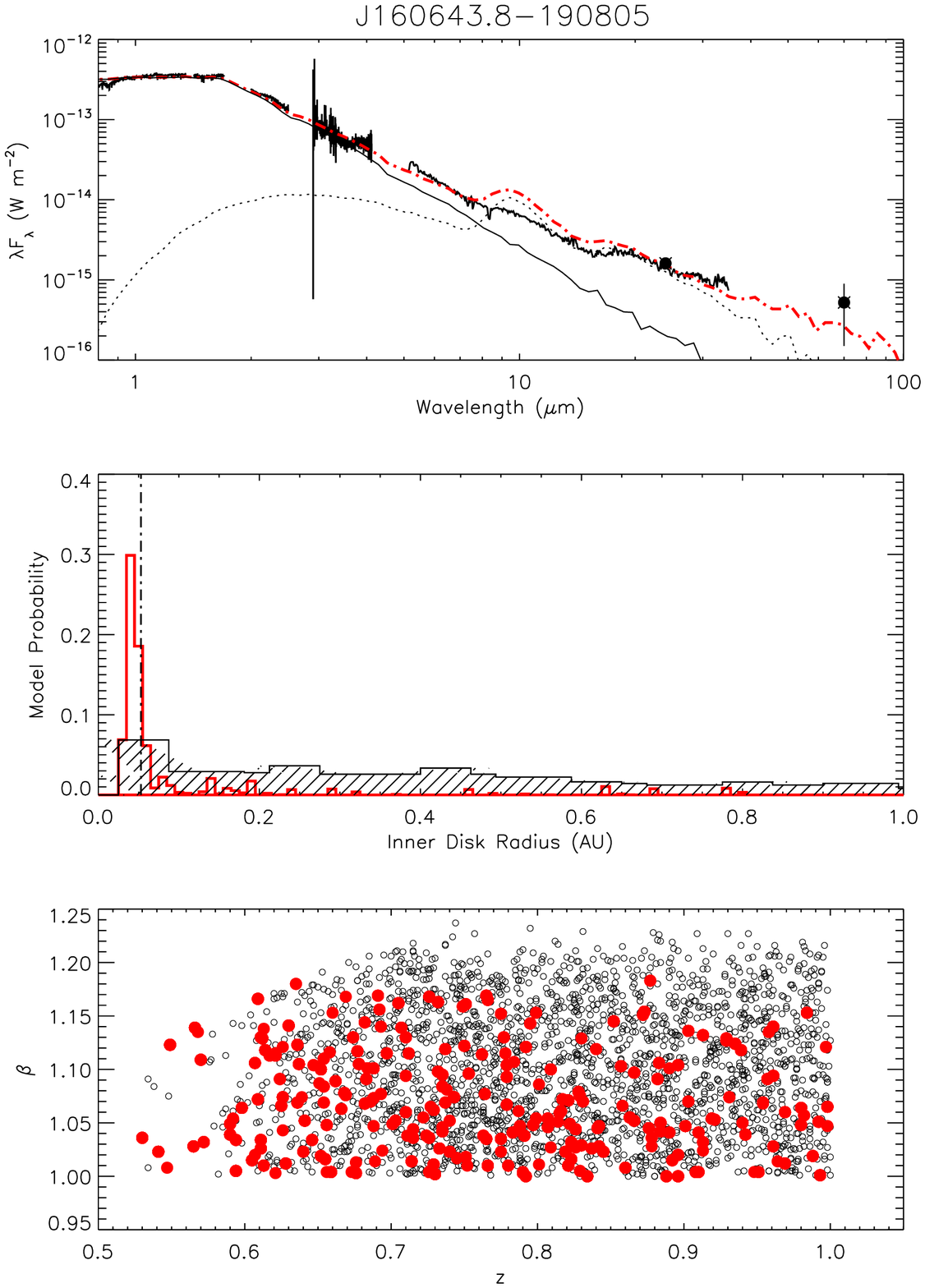}  \vspace{-2cm}    
\caption[f5c.ps]{(continued)
\label{f5c}}
\end{figure}
\clearpage

\addtocounter{figure}{-1}
\addtocounter{subfigure}{1}
\clearpage
\begin{figure}
\epsscale{0.5}
\hspace{2cm}  \vspace{2cm}  \includegraphics[width=11cm,angle=0]{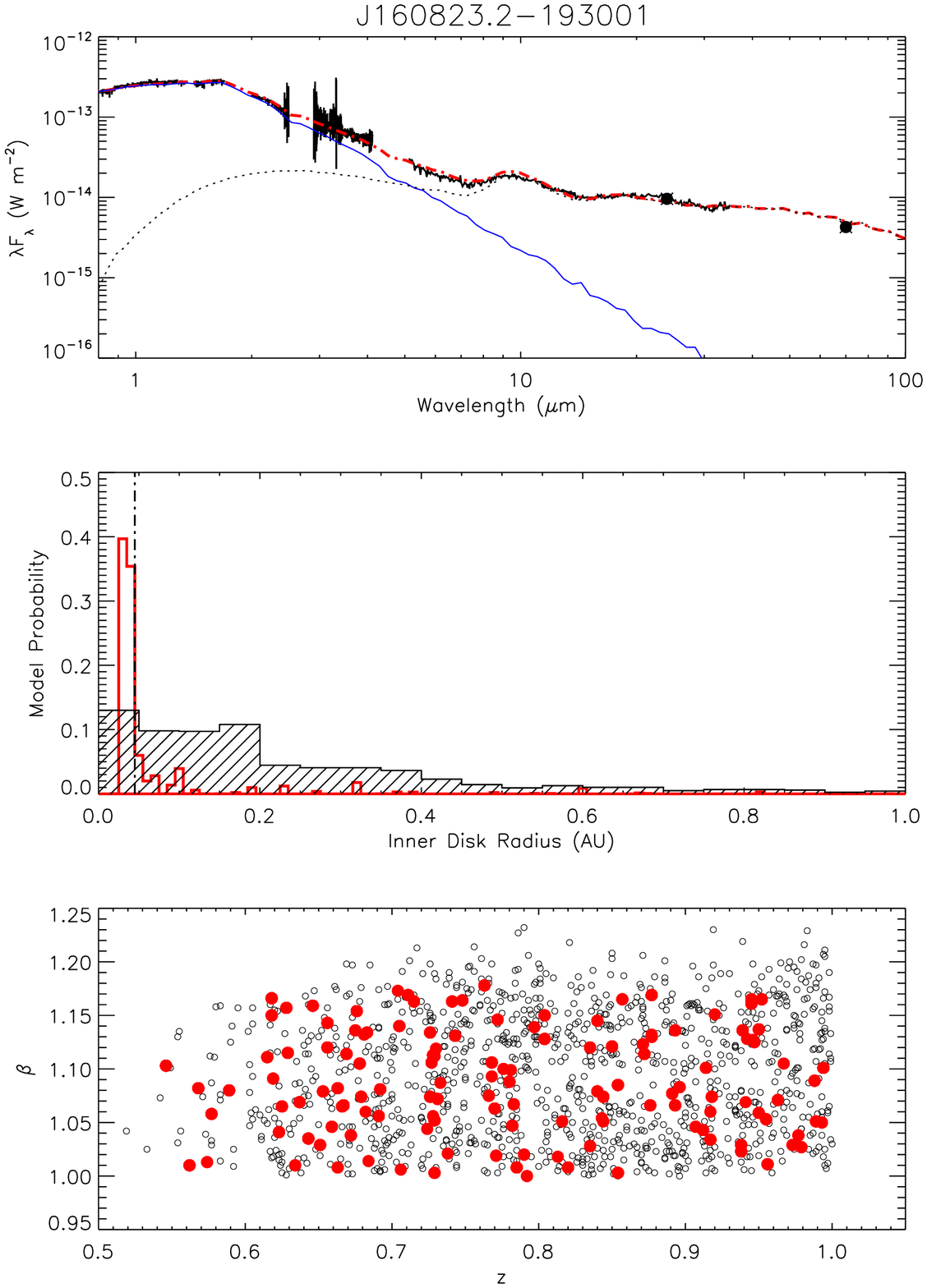}  \vspace{-2cm}    
\caption[f5d.ps]{(continued)
\label{f5d}}
\end{figure}
\clearpage

\addtocounter{figure}{-1}
\addtocounter{subfigure}{1}
\clearpage
\begin{figure}
\epsscale{0.5}
\hspace{2cm}  \vspace{2cm}  \includegraphics[width=11cm,angle=0]{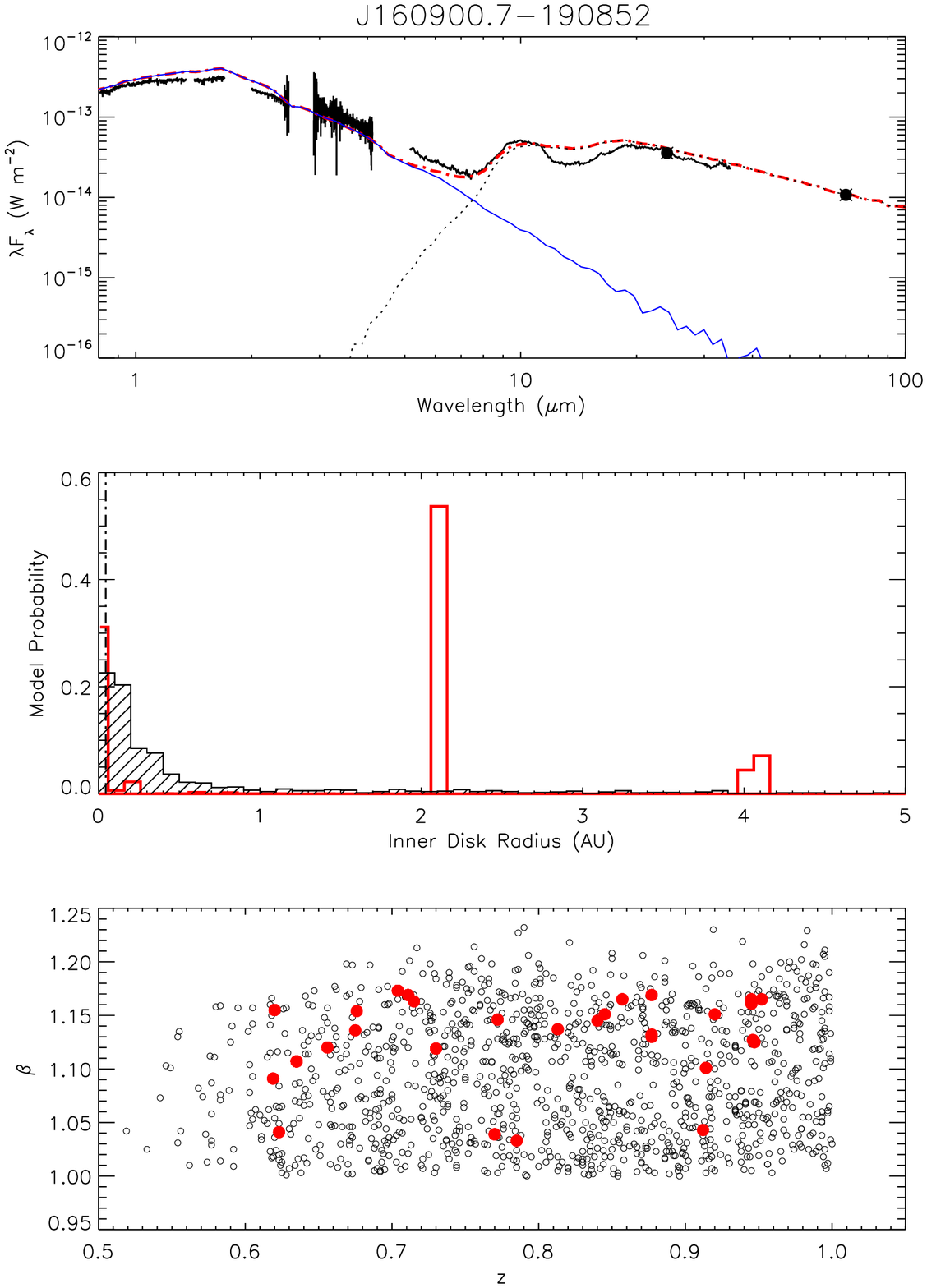}  \vspace{-2cm}    
\caption[f5e.ps]{(continued)
\label{f5e}}
\end{figure}
\clearpage

\addtocounter{figure}{-1}
\addtocounter{subfigure}{1}
\clearpage
\begin{figure}
\epsscale{0.5}
\hspace{2cm}  \vspace{2cm}  \includegraphics[width=11cm,angle=0]{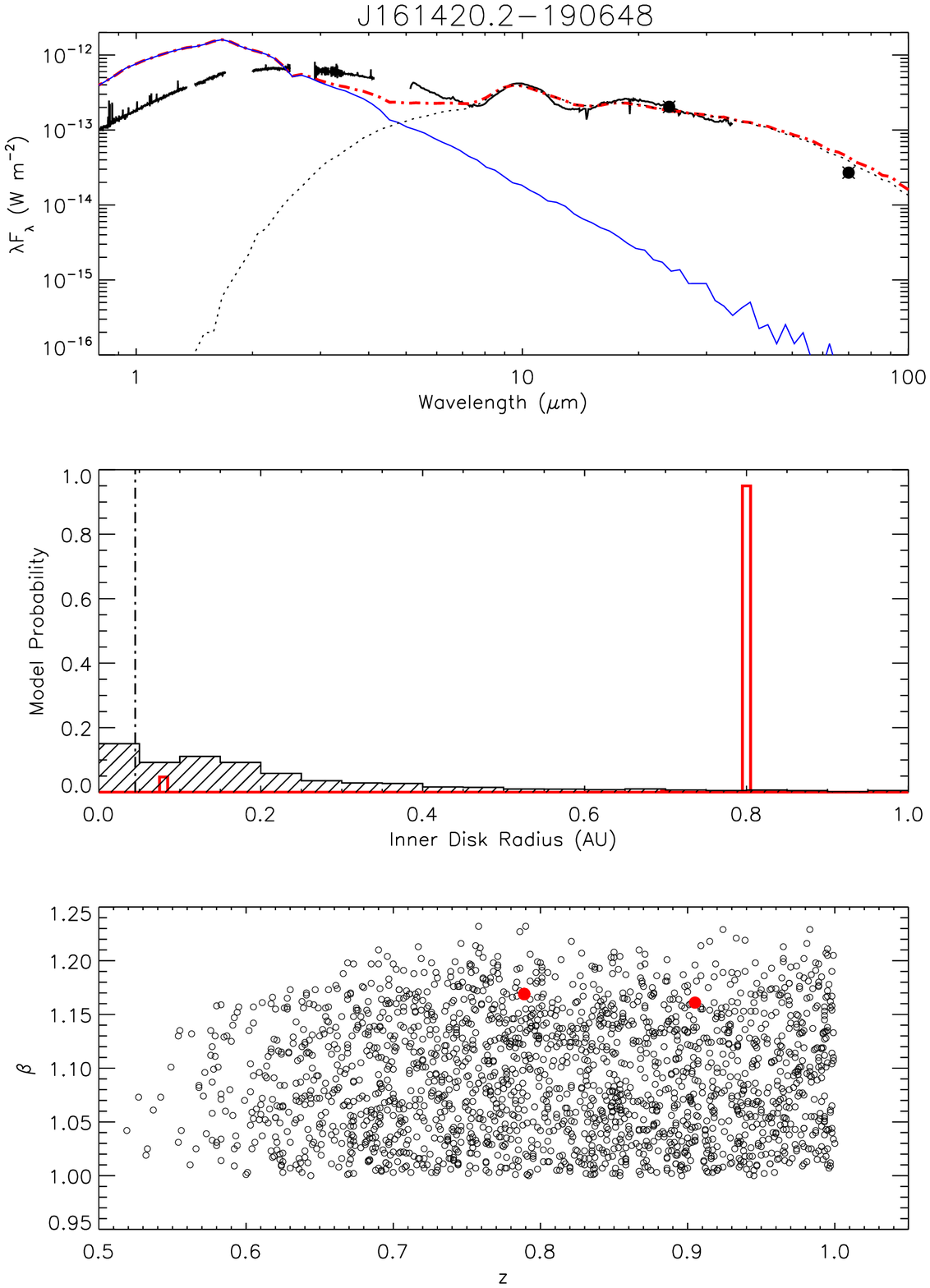}  \vspace{-2cm}    
\caption[f5f.ps]{(continued)
\label{f5f}}
\end{figure}
\clearpage

\addtocounter{figure}{-1}
\addtocounter{subfigure}{1}
\clearpage
\begin{figure}
\epsscale{0.5}
\hspace{2cm}  \vspace{2cm}  \includegraphics[width=11cm,angle=0]{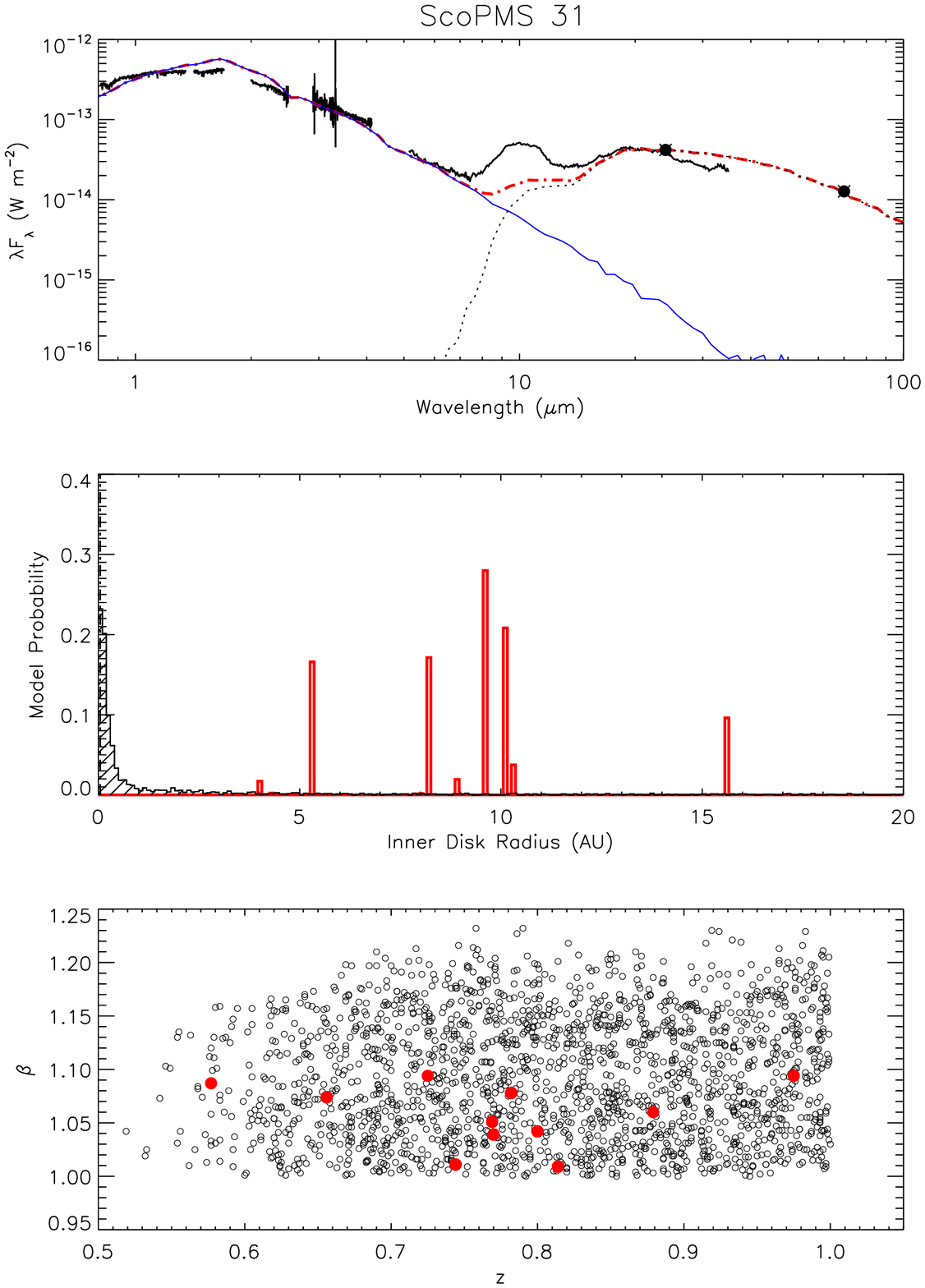}  \vspace{-2cm}    
\caption[f5g.ps]{(continued)
\label{f5g}}
\end{figure}
\clearpage

\addtocounter{figure}{-1}
\addtocounter{subfigure}{1}
\clearpage
\begin{figure}
\epsscale{0.5}
\hspace{2cm}  \vspace{2cm}  \includegraphics[width=11cm,angle=0]{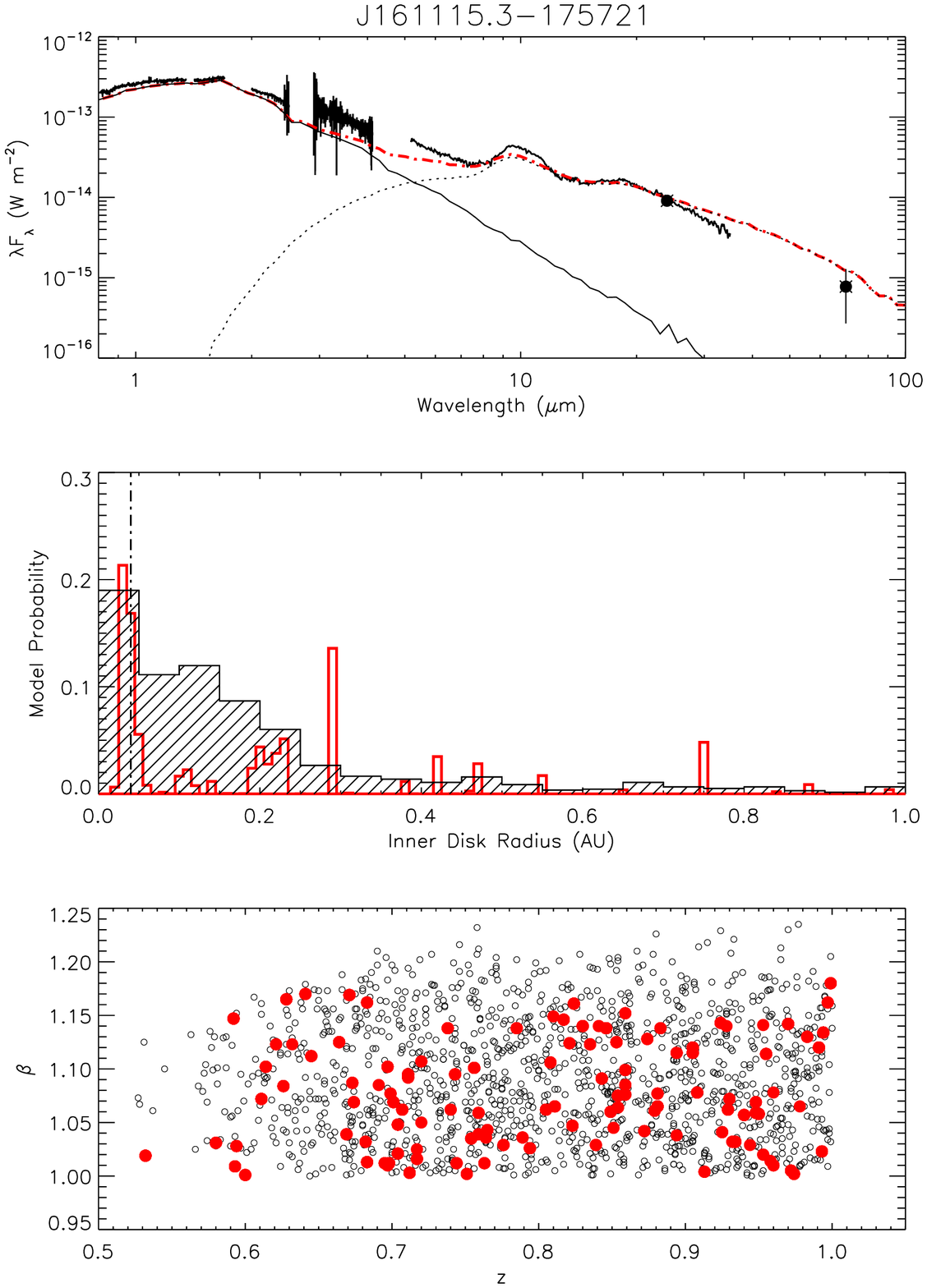}  \vspace{-2cm}    
\caption[f5h.ps]{(continued)
\label{f5h}}
\end{figure}
\clearpage

\addtocounter{figure}{-1}
\addtocounter{subfigure}{1}
\clearpage
\begin{figure}
\epsscale{0.5}
\hspace{2cm}  \vspace{2cm}  \includegraphics[width=11cm,angle=0]{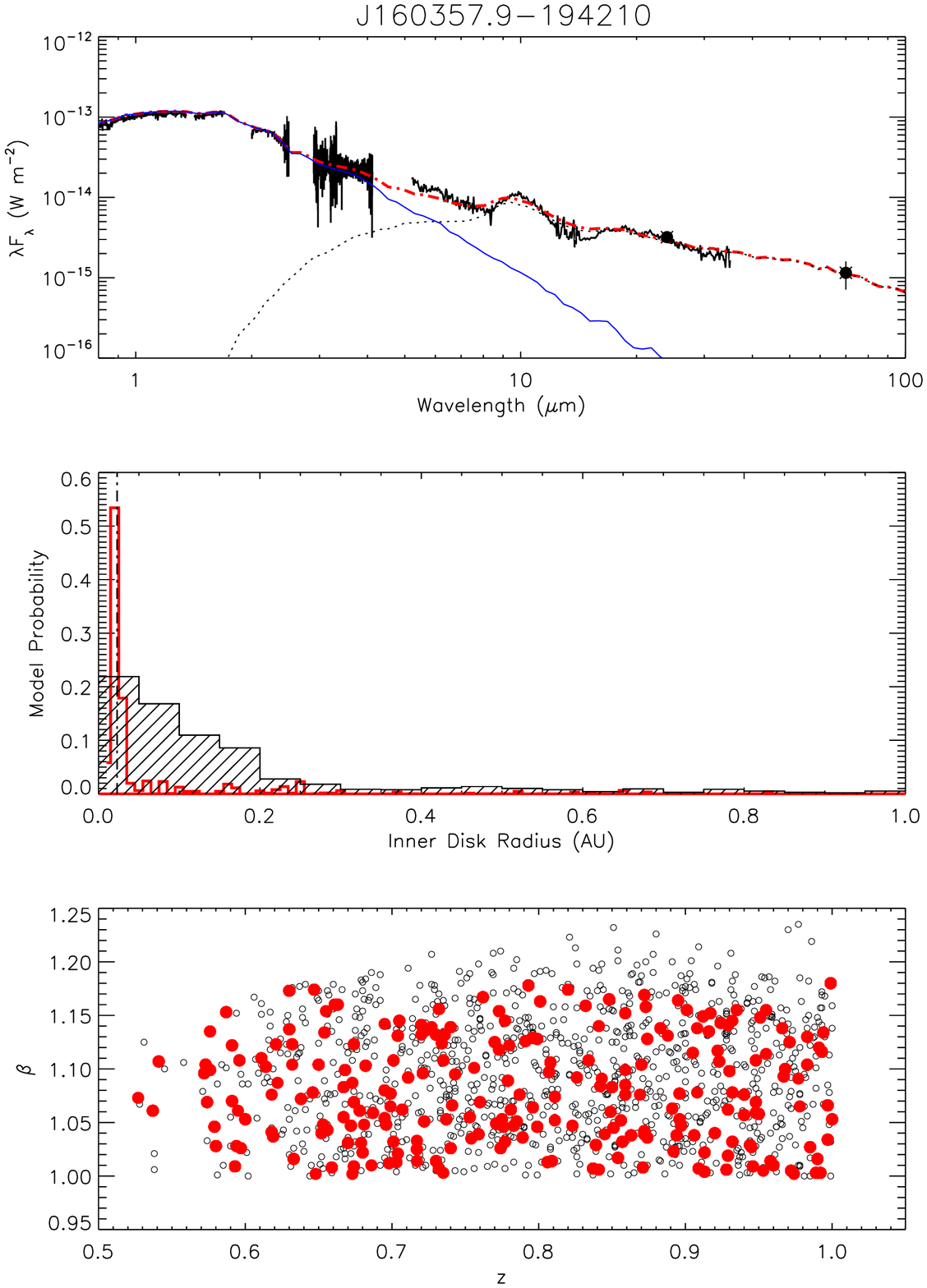}  \vspace{-2cm}    
\caption[f5i.ps]{(continued)
\label{f5i}}
\end{figure}
\clearpage

\begin{deluxetable}{lccccccc}
\tabletypesize{\small}
\tablenum{1}
\tablewidth{0pt}
\rotate
\tablecaption{Stellar Characteristics and Excess Properties of the Upper Scorpius Sample}
\tablehead{
\colhead{Source\tablenotemark{a}}  &  \colhead{SpT\tablenotemark{b}} & \colhead{$A_{V}$\tablenotemark{c}} &  \colhead{Mass\tablenotemark{d}} & \colhead{Radius\tablenotemark{d}} & \colhead{Luminosity\tablenotemark{d}} & \colhead{T$_{eff}$\tablenotemark{d}} & \colhead{T$_{D}$\tablenotemark{e}} \\
       &     &   (mag) & (M$_{\odot}$) & (R$_{\odot}$)  & (L$_{\odot}$) & (K) & (K) \\
}
\startdata
$[$PZ99$]$J161411.0-230536 & K0 & 2.4 & 1.98 & 2.65 & 5.62 & 5329 & 700  \\
$[$PZ99$]$J160421.7-213028 & K2 & 1.0 & 1.08 & 1.14 & 0.74 & 4893 & 900  \\
$[$PZ99$]$J160357.6-203105 & K5 & 0.9 & 1.09 & 1.59 & 0.82 & 4321 & 1300 \\
J160643.8-190805           & K6 & 1.9 & 0.95 & 1.42 & 0.65 & 4205 & $\le$500 \\ 
J160823.2-193001           & K9 & 1.5 & 0.70 & 1.23 & 0.37 & 3963 & 800  \\
J160900.7-190852           & K9 & 0.8 & 0.69 & 1.24 & 0.43 & 3963 & 1300 \\
J161420.2-190648           & M0 & 1.8 & 0.56 & 1.55 & 0.52 & 3840 & 1300 \\
ScoPMS 31                  & M0.5& 0.9& 0.52 & 1.63 & 0.60 & 3782 & 900 \\
J161115.3-175721           & M1 & 1.6 & 0.48 & 1.55 & 0.48 & 3725 & 900 \\
J160357.9-194210           & M2 & 1.7 & 0.40 & 1.06 & 0.17 & 3611 & $\le$500 \\
J160545.4-202308           & M2 & 2.2 & 0.36 & 1.05 & 0.16 & 3530 & 800 \\
J155829.8-231007           & M3 & 1.3 & 0.25 & 0.60 & 0.05 & 3380 & 1300 \\
\enddata
\tablenotetext{a}{Identifiers from Carpenter et al. (2006).}
\tablenotetext{b}{Spectral type from the literature.}
\tablenotetext{c}{Extinction estimates are taken from Preibisch \& Zinnecker (1999) and Preibisch et al. (2002).}
\tablenotetext{d}{From the models of Siess et al. (2000), assuming a distance of 145 pc.}
\tablenotetext{e}{Dust temperature derived from the blackbody fits of the continuum excess emission.}
\end{deluxetable}

\begin{deluxetable}{lcccccc}
\tabletypesize{\small}
\tablenum{2}
\tablewidth{0pt}
\rotate
\tablecaption{Accretion Luminosities and Derived Mass Accretion Rates}
\tablehead{
\colhead{Source}  &   \colhead{EW(Pa$\beta$)\tablenotemark{a}}  & \colhead{EW(Br$\gamma$)\tablenotemark{a}}  & \colhead{log $L_{acc}$(Pa$\beta$)/$L_{\odot}$ \tablenotemark{b}} & \colhead{log $L_{acc}$(Br$\gamma$)/$L_{\odot}$ \tablenotemark{b}} & \colhead{log $\dot{M}$(Pa$\beta$)\tablenotemark{c}} & \colhead{log $\dot{M}$(Br$\gamma$)\tablenotemark{c}} \\
                           & (\AA) &  (\AA)  &  &  &    &    \\
}
\startdata
$[$PZ99$]$J160357.6-203105 & $-$0.51 & $-$0.58: & $-$2.23$\pm$0.18  & $-$2.00$\pm$0.23 & $-$9.47$\pm$0.18 & $-$9.23$\pm$0.27 \\
J160900.7-190852           & $-$1.35 & $-$2.20  & $-$2.04$\pm$0.15  & $-$1.74$\pm$0.18 & $-$9.19$\pm$0.15 & $-$8.89$\pm$0.18 \\
J161420.2-190648           & $-$3.40 & $-$3.69  & $-$1.48$\pm$0.10  & $-$0.65$\pm$0.10 & $-$8.40$\pm$0.10 & $-$7.60$\pm$0.10 \\
ScoPMS 31                  &  ...    & $-$1.45  & ...               & $-$1.62$\pm$0.17 & ...              & $-$8.52$\pm$0.17 \\
J155829.8-231007           & $-$4.15 & $-$3.27  & $-$2.37$\pm$0.20  & $-$2.50$\pm$0.31 & $-$9.39$\pm$0.20 & $-$9.52$\pm$0.31 \\
\enddata
\tablenotetext{a}{Negative equivalent width implies emission.}
\tablenotetext{b}{Accretion luminosity determined using the linear relationship of Muzerolle et al. (1998).}
\tablenotetext{c}{$\dot{M}$ derived assuming R$_{*}$, T$_{eff}$, and M$_{*}$ values listed in Table 1.}
\end{deluxetable}

\begin{deluxetable}{lccccccc}
\tabletypesize{\small}
\tablenum{3}
\tablewidth{0pt}
\rotate
\tablecaption{Accretion Disk Models} 
\tablehead{
\colhead{Source} & \colhead{T$_{eff}$ Range\tablenotemark{a}} & \colhead{Number\tablenotemark{b}} &  \colhead{Model ID\tablenotemark{c}} & \colhead{A$_{V}$\tablenotemark{d}} & \colhead{d\tablenotemark{d}} & \colhead{$i$\tablenotemark{d}} &  \colhead{$\chi^{2}$\tablenotemark{d}} \\
                 &       (K)                 &                  &                     &   (mag)           &  (pc)       &  ($^{\circ}$) & \\
}
\startdata
$[$PZ99$]$J161411.0-230536 & 4900--5410 & 500 & 3014849 & 1.0 & 125 & 75.5 & 6.30 \\
$[$PZ99$]$J160357.6-203105 & 4205--4590 & 2993 & 3012326 & 2.0 & 125 & 31.8 & 13.44 \\
J160643.8-190805           & 4060--4350 & 2753 & 3005191 & 0.5 & 185 & 75.5 & 1.51 \\
J160823.2-193001           & 4850--4060 & 1190 & 3013325 & 0.75 & 165 & 41.4 & 0.34 \\
J160900.7-190852           & 4850--4060 & 1190 & 3019185 & 1.0 & 145 & 81.4 & 4.72 \\
J161420.2-190648           & 3580--4060 & 1998 & 3008376 & 3.0 & 125 & 56.6 & 58.91 \\
ScoPMS 31                  & 3729--4060 & 1917 & 3002397 & 2.0 & 165 & 81.4 & 2.19 \\
J161115.3-175721           & 3580--3850 & 1409 & 3016046 & 2.0 & 165 & 41.4 & 2.87 \\
J160357.9-194210           & 3470--3720 & 1180 & 3018769 & 0.5 & 155 & 63.3 & 0.42 \\
\enddata
\tablenotetext{a}{T$_{eff}$ range considered for the adopted spectral type of the source.}
\tablenotetext{b}{The number of models having T$_{eff}$ values within the specified range.}
\tablenotetext{c}{The best-fitting model identification number from Robitaille et al. (2006).}
\tablenotetext{d}{Extinction, distance, inclination angle, and reduced $\chi^{2}$ values for the best-fitting model.}
\end{deluxetable}

\begin{deluxetable}{lccccccccccc}
\tabletypesize{\tiny}
\tablenum{4}
\tablewidth{0pt}
\rotate
\tablecaption{Inferred Disk Properties of the Upper Scorpius Sources}
\tablehead{
\colhead{Source}  &  \multicolumn{3}{c}{$M_{disk}$ (M$_{\odot}$)\tablenotemark{a}} & & \multicolumn{3}{c}{$\dot{M}$ (M$_{\odot}$ yr$^{-1}$)\tablenotemark{a}} & & \multicolumn{3}{c}{$R_{in}$ (AU)\tablenotemark{a}} \\
\cline{2-4} \cline{6-8} \cline{10-12} \\
                  & \colhead{(min)} & \colhead{(best)} & \colhead{(max)} & & \colhead{(min)} & \colhead{(best)} & \colhead{(max)} & & \colhead{(min)} & \colhead{(best)} & \colhead{(max)} \\ 
}
\startdata
$[$PZ99$]$J161411.0-230536 &  7.18E-7  & 1.06E-5 &  1.16E-2   & &  2.88E-13  & 2.43E-11 & 5.13E-8  & &  0.10 &  0.87  &  2.24    \\
$[$PZ99$]$J160357.6-203105 &  2.60E-6  & 2.62E-5 &  3.47E-2   & &  6.79E-13  & 1.16E-11 & 6.97E-8  & &  0.06 &  0.07  &  0.14    \\
J160643.8-190805           &  5.30E-8  & 3.70E-6 &  5.34E-3   & &  5.06E-15  & 2.28E-14 & 2.22E-9  & &  0.04 &  0.05  &  10.00   \\
J160823.2-193001           &  6.07E-6  & 4.30E-4 &  1.72E-2   & &  4.44E-12  & 2.02E-11 & 3.25E-8  & &  0.03 &  0.04  &  1.19    \\
J160900.7-190852           &  1.44E-5  & 6.56E-3 &  6.56E-3   & &  4.44E-12  & 1.43E-9 & 6.05E-9   & &  0.04 &  2.18  &  4.13    \\
J161420.2-190648           &  4.28E-5  & 4.28E-5 &  1.36E-3   & &  3.31E-10  & 3.31E-10 & 1.17E-7  & &  0.08 &  0.80  &  0.80    \\
ScoPMS 31                  &  5.90E-5  & 8.24E-4 &  1.45E-3   & &  1.31E-12  & 3.57E-9  & 5.20E-9  & &  4.08 &  9.62  &  15.70    \\
J161115.3-175721           &  3.52E-7  & 3.53E-5 &  1.78E-2   & &  2.69E-13  & 1.74E-11 & 4.10E-8  & &  0.03 &  0.29  &  1.19  \\
J160357.9-194210           &  3.62E-7  & 2.50E-4 &  9.50E-3   & &  2.41E-13  & 2.16E-11 & 1.63E-8  & &  0.02 &  0.17  &  1.29  \\ 
\enddata
\tablenotetext{a}{Minimum, best-fitting, and maximum $M_{disk}$, $\dot{M}$, and $R_{in}$ values of the subset of best-fitting models of Robitaille et al. (2006).}
\end{deluxetable}

\end{document}